   \documentclass[twoside,reqno,11pt]{fcaa-var} %

\usepackage{graphicx}
\usepackage{epsfig}
\usepackage{amsthm}
\usepackage{amsmath}
\usepackage{latexsym}
\usepackage{amsfonts}
\usepackage{amssymb}

\newtheoremstyle{theorem}
  {15pt}          
  {15pt}  
  {\sl}  
  {\parindent}
  {\sc}  
  {. }   
  { }    
  {}     
\theoremstyle{theorem}
\newtheorem{lemma}{Lemma}[section]
\newtheorem{theorem}{Theorem}[section]
\newtheorem{corollary}{Corollary}[section]

\newtheoremstyle{defi}
  {15pt}          
  {15pt}  
  {\rm}  
  {\parindent}     
  {\sc}  
  {. }    
  { }    
  {}     
\theoremstyle{defi}
\newtheorem{definition}{Definition}[section]

 \def\proofname{\indent\rm P\ r\ o\ o\ f.}
 \def\proofend{\hfill$\Box$}
 \def\qed{\hfill$\Box$} 

 \usepackage{hyperref} 

 \def\theequation{\arabic{section}.\arabic{equation}}

 \def\themycopyrightfootnote{\vspace*{3pt}
$^*$Accepted for publication in {\it Fractional Calculus and Applied Analysis}
   \hfill  \vspace*{-36pt}
  }

\usepackage{graphicx}
\usepackage{color}

\def\sign{\hbox{sign}}
\def\e{\hbox{e}}
\def\exp{\hbox{exp}}
\def\q{\quad} 
\def\l{\left} \def\r{\right}

\newcommand{\be}{\begin{equation}}
\newcommand{\ee}{\end{equation}}
\newcommand{\Kabt}{K_{\alpha,\beta}^{\theta}}
\newcommand{\Kab}{K_{\alpha,\beta}^{0}}
\newcommand{\Lat}{L_{\alpha}^{\theta}}
\newcommand{\Mb}{M_{\beta}}
\newcommand{\Xab}{X_{\alpha,\beta}}

\newcommand{\bx}{{\bf{x}}}

\newcommand{\mG}{\mathcal{G}}
\newcommand{\mLe}{\mathcal{L}^{ext}}
\newcommand{\mMb}{\mathcal{M}_\beta}

\title[STOCHASTIC SOLUTION WITH GAUSSIAN \dots]
{A stochastic solution with Gaussian stationary increments of the symmetric space-time fractional diffusion equation$^*$}
\author[\normalsize G. Pagnini, P. Paradisi]{\normalsize Gianni Pagnini $^{1,2}$, Paolo Paradisi $^{1,3}$}

\begin{document}

 \vbox to 2.5cm { \vfill }


 \bigskip \medskip

 \begin{abstract}
The stochastic solution with Gaussian stationary increments is establihsed for the symmetric
space-time fractional diffusion equation when $0 < \beta < \alpha \le 2$,
where $0 < \beta \le 1$ and $0 < \alpha \le 2$ are the fractional derivation orders
in time and space, respectively.
This solution is provided by imposing the identity between two probability
density functions resulting (i) from a new integral representation formula
of the fundamental solution of the symmetric space-time fractional diffusion equation
and (ii) from the product of two independent random variables.
This is an alternative method with respect to previous approaches such as
the scaling limit of the continuos time random walk, the parametric subordination and
the subordinated Langevin equation.
A new integral representation formula for the fundamental solution of the space-time
fractional diffusion equation is firstly derived.
It is then shown that, in the symmetric case, a stochastic solution can be obtained by
a Gaussian process with stationary increments and with a random wideness scale 
variable distributed according to an arrangement of two extremal L\'evy stable densities.
This stochastic solution is self-similar with stationary increments and uniquely
defined in a statistical sense by the mean and the covariance structure.
Numerical simulations are carried out by choosing as Gaussian process the fractional Brownian motion.
Sample paths and probability densities functions are shown to be in agreement with the
fundamental solution of the symmetric space-time fractional diffusion equation.

 \medskip


 \smallskip


 \end{abstract}

 \maketitle

 \vspace*{-16pt}




\section{Introduction}
\label{sec:introduction}
\setcounter{section}{1}
\setcounter{equation}{0}\setcounter{theorem}{0}

{\it Space-time fractional diffusion} was originally introduced in physics by Zaslavsky
\cite{zaslavsky-1992,zaslavsky-c-1994,zaslavsky-pd-1994} to study
chaotic Hamiltonian dynamics in low dimensional systems with the
specific aim to model the so-called {\it anomalous diffusion} (see
also \cite{zaslavsky_etal-pr-1997,zaslavsky-pr-2002,zaslavsky-2005}).
The label {\it anomalous diffusion} is assigned to processes whose
variance does not grow linearly in time
\cite{schmiedeberg_etal-jsmte-2009}, 
in contrast to Gaussian {\it normal diffusion} that is mainly characterized by such linear law.
Anomalous diffusion has been experimentally observed several
times and definitively established in nature not only in chaotic
dynamical systems (see, 
e.g., References \cite{metzler_etal-jpa-2004,ratynskaia_etal-prl-2006,dieterich_etal-pnas-2008, difpradalier_etal-pre-2010,chevrollier_etal-epjd-2010}).

\smallskip
Zaslavsky argued that, since chaotic dynamics is a physical phenomenon
whose evolution bridges between a completely regular integrable system
and a completely random process \cite{zaslavsky-pr-2002}, kinetic
equations and statistical tools arise as modelling methods. 
In these cases the classical diffusion paradigm, which is based on a local
and linear flux-gradient relationship, does not hold, thus a non-local and/or
non-linear relationship is needed.
In the Fractional Calculus approach the idea is to maintain
a linear relationship, while introducing a non-local dependence in the 
flux-gradient relationship by means of integrals with inverse power-law kernels
\cite{paradisi_pceb01, paradisi_pa01}. 
This is in agreement with the observation of non-local behavior
in many natural phenomena, such as the emergence of spatially extended coherent
structures in the turbulent atmospheric boundary layer
\cite{paradisi_pceb01, paradisi_epjst09, paradisi_romp12, paradisi_npg12}.

\smallskip
Fractional Calculus \cite{podlubny-1999,baleanu_etal-2012} is nowadays
recognized to be a useful mathematical tool for modelling such linear non-local effects.
In this framework, non-locality can be considered in time ({\it
  time-fractional diffusion})
\cite{mainardi-csf-1996,compte-pre-1996,schneider_etal-jmp-1989,gorenflo_etal-nd-2002,weron_etal-pre-2008}
or in space ({\it space-fractional diffusion})
\cite{compte-pre-1996,fogedby-pre-1994,honkonen-pre-1996,gorenflo_etal-fcaa-1998},
as well as both in space and time ({\it space-time fractional
  diffusion})
\cite{zaslavsky-2005,fogedby-pre-1994,saichev_etal-c-1997,scalas_etal-pa-2000,mainardi_etal-fcaa-2001,gorenflo_etal-cp-2002,gorenflo_etal-pa-2002,barkai-cp-2002,hughes-pre-2002,metzler_etal-cp-2002,meerschaert_etal-pre-2002,delcastillonegrete-pp-2004,delcastillonegrete_etal-prl-2005,delcastillonegrete-npg-2010,baeumer_etal-jap-2009,dybiec-jsmte-2009}.

\smallskip
Moreover, when there is no separation of time scales between the
microscopic and the macroscopic level of the process, the randomness
of the microscopic level is, at least partially, transmitted to the macroscopic
level and the macroscopic dynamics is correctly described by means of non-local
operators with self-similar features, hence the emergence of
Fractional Calculus \cite{grigolini_etal-pre-1999}.  Further,
fractional integro-differential equations are related to the fractal
properties of phenomena \cite{rocco_etal-pa-1999}.  However, the
fractional kinetics strongly differs from the usual kinetics because
some moments of the probability density function (PDF) of particle
displacement can be infinite and the fluctuations from the equilibrium
state have a broad distribution of relaxation times
\cite{sokolov_etal-pt-2002,zaslavsky-pr-2002,klafter_etal-pw-2005}.

\smallskip
The {\it space-time fractional diffusion} equation reads \cite{mainardi_etal-fcaa-2001} 
\begin{subequations}
\be {_tD_*^\beta} \, u(x;t) =
     {_xD_\theta^\alpha} \, u(x;t) \,, \quad -\infty < x < + \infty
     \,, \quad t \ge 0 \,,
\label{STFDE0} 
\ee
\be
u(x;0)=u_0(x) \,, \quad u(\pm\infty;t)=0 \,.
\label{STFDE0_initcond} 
\ee 
\end{subequations}
where $_tD_*^\beta$ is the {\it Caputo time-fractional derivative}
of order $\beta$ and $_xD_\theta^\alpha$ is the {\it Riesz-Feller
space-fractional derivative} of order $\alpha$ and asymmetry
parameter $\theta$.  
The real parameters $\alpha$, $\theta$ and $\beta$ 
are restricted as follows 
\be 
\hspace{-0.1truecm}
0<\alpha\le 2 \,, \q
|\theta| \le \min \{\alpha, 2-\alpha\} \,, \quad 0<\beta \le 1 \q
\hbox{or} \q 1<\beta\leq \alpha \leq 2 \,.
\label{def:alphabetatheta}
\ee 
The definitions of the fractional differential operators
${_tD_*^\beta}$ and ${_xD_\theta^\alpha}$ with a brief review of
(\ref{STFDE0}) are reported in Appendix (see also
References \cite{mainardi_etal-fcaa-2001,gorenflo_etal-klm-2012}).

\smallskip
The general solution of (\ref{STFDE0}) can be represented as 
\be
u(x;t)=\int_{-\infty}^{+\infty} \Kabt(x-\xi;t) \, u_0(\xi) \, d\xi \,, 
\ee 
where $\Kabt(x;t)$ is the fundamental solution, or Green function, which is 
obtained by setting in (\ref{STFDE0_initcond}) the initial condition
$u_0(x)=\delta(x)$.
%
%
Particular cases of Eq. (\ref{STFDE0}) are the space-fractional diffusion 
equation when $\beta=1$, the time-fractional
diffusion equation when $\alpha=2$ and the (Gaussian) parabolic diffusion equation when $\alpha=2$
and $\beta=1$.  Furthermore when $\alpha=\beta=2$ the D'Alembert wave
equation is recovered.

\smallskip
Space-time fractional diffusion equation (\ref{STFDE0}) was
analytically considered by many authors
\cite{saichev_etal-c-1997,uchaikin-ijtp-2000,gorenflo_etal-fcaa-2000,pagnini-tesi-2000,metzler_etal-jpa-2004,mainardi_etal-fcaa-2001}.
The fundamental solution has been expressed by the Mellin--Barnes integral representation (\ref{MB})
\cite{pagnini-tesi-2000,mainardi_etal-fcaa-2001} as well as in terms
of H-Fox function \cite{pagnini-tesi-2000,mainardi_etal-jcam-2005}.

Solutions of equation (\ref{STFDE0}) have been recognized to be good fitting in
anomalous diffusion processes such as, for example, non-diffusive chaotic
transport by Rossby waves in zonal flow \cite{delcastillonegrete-npg-2010}, 
transport in pressure-gradient-driven plasma turbulence
\cite{delcastillonegrete-npg-2010,delcastillonegrete_etal-prl-2005,delcastillonegrete-pp-2004}, transport with perturbative effects in magnetically confined 
fusion plasmas \cite{delcastillonegrete_etal-nf-2008}, non-diffusive tracer
transport in a zonal flow under the effects of finite Larmor radius
\cite{gustafson_etal-pp-2008} or transport in point vortex flow
\cite{leoncini_etal-csf-2004}.  

\smallskip
On the physical ground, the
time-fractional derivative is related to the non-Markovianity, thus long-range
memory, and the space-fractional derivative to non-Gaussian particle 
displacement PDF with heavy tails, thus non-locality. In particular, in the 
Rossby waves problem, the
trapping effect of the vortices gives rise to non-Markovian effects, 
determining the emergence of subdiffusive behavior, 
and the zonal shear flows give rise to non-Gaussian particle
displacement \cite{delcastillonegrete-npg-2010}.  In the plasma
physics problem, the non-Markovian effects are due to the trapping in
electrostatic eddies and the non-Gaussian particle displacements
result from avalanche-like radial relaxation events
\cite{delcastillonegrete-pp-2004,delcastillonegrete_etal-prl-2005,delcastillonegrete-npg-2010}.
In the context of flows in porous media, fractional time derivatives
describes particles that remain motionless for extended periods of
time while fractional space derivatives model large motions through
highly conductive layers or fractures
\cite{meerschaert_etal-pre-2002,benson_etal-awr-2013}.

Castiglione {\it et al.} \cite{castiglione_etal-pd-1999} have highlighted
that fractional diffusion equation (\ref{STFDE0}) fails to model
{\it strong anomalous diffusion},
namely those diffusive processes with the scaling laws of statistical moments depending 
on the order of the moment under consideration: $\langle X^m \rangle(t) \sim t^m g(m)$, 
where $g(m)$ is a function of the moment order $m$. 
This is typically associated with multi-scaling or multi-fractal signals. 
However, in many systems the fractional diffusion approximation was shown to be 
still valid, and this is even more true when considering the long-time behavior
\cite{schmiedeberg_etal-jsmte-2009, barkai-cp-2002}.

\smallskip
Since (\ref{STFDE0}) can be understood as a Master equation and its
solution as a PDF, the formulation of
the underlying {\it stochastic process} is important to physically depict
the anomalous diffusion at the micro/mesoscopic scale.  
In this respect, it is well-known that the classical Gaussian diffusion,
namely the {\it Brownian motion} (Bm), is stochastically described by the Wiener process.
In this paper a {\it stochastic solution} for the {\it symmetric space-time 
fractional diffusion} equation is derived, where stochastic solution means a
stochastic process whose one-point one-time PDF is the solution of the Master (diffusion) equation. 
The symmetric space-time fractional diffusion equation occurs when $\theta=0$:  
\be {
_tD_*^\beta} \, u(x;t) = {_xD_0^\alpha} \,
u(x;t) = \frac{\partial^\alpha u}{\partial |x|^\alpha} \,, \quad
-\infty < x < + \infty \,, \quad t \ge 0 \,,
\label{STFDE} 
\ee 
with $\alpha$ and $\beta$ ranging as in (\ref{def:alphabetatheta}).  
Preliminary analytic solutions of the symmetric case (\ref{STFDE}) were computed by 
Saichev \& Zaslavsky \cite{saichev_etal-c-1997} and Gorenflo, Iskenderov 
\& Luchko \cite{gorenflo_etal-fcaa-2000}.

\smallskip
It is worth noting that the solution of (\ref{STFDE}) with the proper initial and boundary conditions is unique, 
but this uniqueness is not met in general in the underlying stochastic process. 
In fact, 
the Master equation determines only the one-point one-time PDF, thus there is an
infinite number of stochastic solutions which define $n$-points $n$-times PDFs,
each one characterized by different space-time correlation properties, 
that share the same one-point one-time PDF.   

In this respect it is reminded that a stochastic solution of the Master equation (\ref{STFDE}) is, 
for example, the Continuous Time Random Walk (CTRW) \cite{montroll1964,montroll_etal-jmp-1965,weiss1983}.
The CTRW is a random walk model with crucial events occurring randomly in time, 
thus characterized not only by the probability distribution of the jumps, but
also by the probability distribution of the inter-event times. In general, the events can be
correlated and so the inter-event times. However, the most frequent and
reasonable assumption is that of statistical independence among inter-event times,
thus defining a renewal point process \cite{cox_1962}. 
Interestingly, the renewal property, which has been verified in many real
phenomena \cite{paradisi_epjst09, paradisi_npg12,bianco_cpl07, paradisi_aipcp13, allegrini_csf13}, 
is associated with the emergence of metastable self-organized states, or 
coherent structures, and, in more detail, with an intermittent birth-death 
process of self-organization 
\cite{paradisi_npg12,paradisi_romp12,paradisi_cejp09,akin_pa06,akin_jsmte09, paradisi_bmcsb15,paradisi_caim15}. 
Thus, even if a rigorous derivation of fractional operators
from intermittent self-organization does not exist, fractional diffusion 
equations are expected to emerge in these cases.

The CTRW is an exact stochastic solution when the 
cumulative distribution of the inter-event times is a Mittag--Leffler function 
\cite{scalas_etal-pre-2004,fulger_etal-pre-2008,germano_etal-pre-2009}.
In general, CTRWs with proper scaling limits in the inter-event time
and/or jump distribution satisfy the space and/or time fractional diffusion
equation, but only in the long-time limit
\cite{gorenflo_etal-jcam-2009,gorenflo_etal-epjst-2011,fulger_etal-fcaa-2013}.
Other stochastic solutions were derived by means of the parametric subordination
\cite{gorenflo_etal-jcam-2009,gorenflo_etal-epjst-2011,gorenflo_etal-klm-2012}
and of the subordinated Langevin equation
\cite{fogedby-pre-1994,weron_etal-pre-2008,magdziarz_etal-prl-2008}.
However, all these methods, which are also interconnected, do not have
stationary increments.

\smallskip
Moreover, 
we highlight here also that in certain cases the knowledge of the one-point one-time PDF is not enough
to uniquely infer information on the system, 
as for example the information on the first passage time when geometric constraints are applied 
\cite{sokolov_etal-jpa-2004,meroz_etal-prl-2011,schulz_etal-jpa-2013}.

In this paper we propose an approach that leads to a unique stochastic
solution in the sense that it is fully statistically characterised and has Gaussian stationary increments. 
This stochastic process is essentially associated with a {\it Gaussian stochastic process}
and then fully characterized by its first and second moments. 
In particular we define and characterise a process whose one-point one-time PDF is the solution
of the symmetric space-time fractional diffusion equation (\ref{STFDE}).

Such process is defined as the product of a Gaussian process by an independent and
constant non-negative random variable distributed according to a combination of L\'evy stable densities.
The chosen Gaussian process may have an arbitrary temporal correlation, 
that however must be in agreement with the time dependent variance of the resulting final process.

The choice of this Gaussian process can be made to meet some physical costraints, 
as for example the geometrical constraints mentioned above 
\cite{sokolov_etal-jpa-2004,meroz_etal-prl-2011,schulz_etal-jpa-2013},
and after this some conditions for the statistical characterisation can be derived.

Here, the {\it fractional Brownian motion} (fBm) is chosen, 
among the infinite different Gaussian processes that can be used,
because this choice provides a self-similar process with Gaussian stationary increments,
beside the fact that it is a simple approach for trajectory simulations.

\smallskip
To achieve this stochastic solution with stationary increments
we firstly derive a new {\it integral representation formula} for the fundamental solution of the 
asymmetric space-time fractional diffusion equation (\ref{STFDE0}).
Then, we consider the symmetric case corresponding to Eq. (\ref{STFDE})
and we exploit the identity between the PDFs resulting, on one side, from such new
integral representation formula and, on the other side, from the product of two
independent random variables. 
Finally we establish the correspondence of such product of two independent random variables
with the product of a Gaussian process, e.g., the fBm, by an appropriate random variable.

\smallskip
The rest of the paper is organized as follows. In Section
\ref{sec:subordination} we derive a new  
integral representation formula
for the fundamental solution of the space-time fractional diffusion equation (\ref{STFDE0}).
In Section
\ref{sec:hsssi} we obtain the stochastic solution of (\ref{STFDE}) 
by using the identity between the resulting PDF of the integral representation
formula 
%
%
and the PDF of the product of two independent variables.
In Section \ref{sec:simulations} we show the results of numerical simulations
and, finally, we draw some discussions and conclusions in Section 
\ref{sec:conclusion}. In Appendix we summarize the main properties of 
space-time fractional diffusion equation (\ref{STFDE0}).

\section{Integral representation formulae for the fundamental solution of 
the space-time fractional diffusion equation}
\label{sec:subordination}
\setcounter{section}{2}
\setcounter{equation}{0}\setcounter{theorem}{0}

\subsection{Brief review on integral representation formulae}

Let $Y(\tau)$, $\tau > 0$, be a stochastic process. If the parameter
$\tau$ is randomized according to a second stochastic process $T$ with
non-negative increments, i.e., $\tau=T(t)$, then the resulting process
$X(t)=Y(T(t))$ is said to be {\it subordinated} to $Y(\tau)$, which is
called the {\it parent process}, and to be directed by $T(t)$, which
is the {\it directing process} \cite{feller-1971}.  In diffusive
processes, the parameter $\tau$ is a time-like variable and it is
referred to as the {\it operational time} \cite{dybiec_etal-c-2010}.
In terms of PDFs, such subordination process is embodied by the
following integral formula 
\be 
p(x;t)=\int_0^\infty \psi(x;\tau) \,
\varphi(\tau;t) \, d \tau \,,
\label{subordination0}
\ee
where $p(x;t)$ is the PDF of the resulting process $X(t)$, 
$\psi(x;\tau)$ is the PDF of the parent process $Y(\tau)$ 
and $\varphi(\tau;t)$ the PDF of the directing process $T(t)$.  

Formula
(\ref{subordination0}) is a particular case of {\it integral representation
formula} that can be studied in the framework of the Mellin
transform theory and interpreted as a convolution integral
\cite{mainardi_etal-fcaa-2003,mainardi_etal-jms-2006}.
Many integral representation formulae, which are summarized in the following, 
were studied independently from their stochastic interpretation, which is not 
given only by the subordination approach.

\smallskip
The PDFs under consideration display a self-similar or scaling property, 
i.e., they can be reduced to functions of a similarity variable namely 
$F(x;t)=1/t^\Lambda F_0(x/t^\Lambda)$, where $\Lambda$ is the scaling parameter.
In order to simplify the mathematical notation, and without loss of generality, 
hereinafter the same symbol is used for both the two-variable function $F(x;t)$ 
and the one-variable function $F_0(z)$, $z=x/t^\Lambda$, 
so that the above relationship turns out to be written as 
$F(x;t)=1/t^\Lambda F(x/t^\Lambda)$. 
The chosen notation is not ambiguous as the meaning clearly follows from the 
number of independent variables.

\smallskip
The following valuable integral representation formula for $\Kabt(x;t)$ was
derived by Uchaikin \& Zolotarev
\cite{uchaikin_etal-1999,uchaikin-ijtp-2000} \be \Kabt(x;t) =
\int_0^\infty \Lat(x; (t/y)^\beta) \, L_{\beta}^{-\beta}(y) \, d y
\,,
\label{uchaikin}
\ee 
where 
$\Lat(x;t)$ is the L\'evy stable distribution with scaling parameter 
$\alpha$ and asymmetry parameter $\theta$.
In particular, the L\'evy stable distribution
is the fundamental solution of the space-fractional diffusion equation
that is obtained as special case of (\ref{STFDE0}) when $\beta=1$, i.e.,
$\Lat(x;t)=K_{\alpha,1}^\theta(x;t)$ (see Appendix).
By putting $t/y = \xi^{1/\beta}$, formula (\ref{uchaikin}) becomes
\cite{mainardi_etal-fcaa-2001} 
\be \Kabt(x;t) = \int_0^\infty \Lat
(x;\xi) \, L_{\beta}^{-\beta}(t;\xi) \, \frac{t}{\beta \, \xi} \,
d\xi \,.
\label{m3}
\ee 

This formula is also used in the elegant {\it parametric subordination}
approach, developed by Gorenflo and co-authors 
\cite{gorenflo_etal-csf-2007,gorenflo_etal-jcam-2009,gorenflo_etal-epjst-2011,gorenflo_etal-klm-2012},
and it is based on a systematic and consequent application of the CTRW
integral equation to the various processes involved.
This approach
considers the parent process $Y(\tau)$ and the random walk
$t=T^{-1}(\tau)$, which is the inverse of $\tau=T(t)$ and it is called
the {\it leading process}.  Hence, after the identification of the
particle trajectory with the parent process, from the system composed
by $X=Y(\tau)$ and $t=t(\tau)$ the dummy variable $\tau$ can be
eliminated and the evolution of $X(t)$ obtained.
This is similar to the set of subordinated Langevin equations
proposed by Fogedby \cite{fogedby-pre-1994}:
$$
\frac{dX}{d\tau} = \eta(\tau) \,, \quad 
\frac{dt}{d\tau} = \xi(\tau) \,, 
$$ 
where $\eta(\tau)$
and $\xi(\tau)$ are independent noises whose distributions are related
to the parent and the leading process, respectively (see also
References \cite{kleinhans_etal-pre-2007,gorenflo_etal-csf-2007,weron_etal-pre-2008,magdziarz_etal-prl-2008,eule_etal-epl-2009,dybiec_etal-c-2010}).

\smallskip
Integral representation formulae (\ref{uchaikin}) and (\ref{m3}) were used also
in a theoretical statistical approach to study space-time fractional
diffusion
\cite{baeumer_etal-fcaa-2001,baeumer_etal-jap-2009,baeumer_etal-tams-2009,meerschaert_etal-2012}
as well as to derive the stochastic solution of (\ref{STFDE0})
\cite{baeumer_etal-fcaa-2001,meerschaert_etal-pre-2002}.

\smallskip
Further important integral representation formulae were derived in literature
\cite{mainardi_etal-fcaa-2001,mainardi_etal-fcaa-2003}. In
particular, starting from \cite[Eq. (6.16)]{mainardi_etal-fcaa-2001},
it results:
\begin{subequations}
\be 
\Kabt(z)= \alpha \,
\int_0^\infty \xi ^{\alpha-1} \, M_{\beta}\left(\xi^{\alpha}\right) \,
\Lat(z/\xi) \, \frac{d\xi}{\xi} \,, \quad 0< \beta \le 1 \,,
\label{lumapa1}
\ee 
\be 
\Kabt(z)= \int_0^{\infty} M_{\beta/\alpha}(\xi) \,
N_\alpha^\theta(z/\xi) \, \frac{d\xi}{\xi} \,, \quad 0<
\beta/\alpha \le 1 \,,
\label{lumapa2}
\ee 
\end{subequations}
where $M_{\nu}(\xi)$, $0 < \nu < 1$, is the M-Wright/Mainardi function 
which is also related to the fundamental solution of the time-fractional diffusion equation
that is obtained as special case of (\ref{STFDE0}) when $\alpha=2$,
i.e., $M_{\beta/2}(x;t)/2=K_{2,\beta}^0(x;t)$,
and $N_\alpha^\theta(x;t)$ is the fundamental solution of the neutral fractional diffusion equation
that is obtained as special case of (\ref{STFDE0}) when $0 < \alpha=\beta<2$,
i.e., $N_\alpha^\theta(x;t)=K_{\alpha,\alpha}^\theta(x;t)$
(see Appendix for more information on $M_{\beta}$, $\Lat$ and $N_\alpha^\theta$).

By replacing $z$ with $x/t^{\beta/\alpha}$, after the changes of variable 
$\xi=\tau^{1/\alpha}/t^{\beta/\alpha}$ and $\xi=\tau/t^{\beta/\alpha}$ in 
(\ref{lumapa1}) and (\ref{lumapa2}), respectively, it follows
\cite{mainardi_etal-fcaa-2001,mainardi_etal-fcaa-2003}: 
\begin{subequations}
\be
\hspace{-0.6truecm}
t^{-\beta/\alpha} \Kabt\l(\frac{x}{t^{\beta /\alpha}}\r)
\!=\! \int_0^{\infty} 
\! \Lat\l(\frac{x}{\tau^{1/\alpha}}\r) t^{-\beta} 
M_\beta\l(\frac{\tau}{^\beta}\r) 
\frac{d\tau}{\tau^{1/\alpha}} \,, 
\,\, 0 < \beta \le 1 \,, \!\!\!\!\!
\label{subordinationstfde0}
\ee 
\be 
\hspace{-0.6truecm}
t^{-\beta/\alpha} 
\Kabt\l(\frac{x}{t^{\beta/\alpha}}\r) \!=\!\! \int_0^{\infty}  
\!\! t^{-\beta/\alpha} 
M_{\beta/\alpha}\l(\frac{\tau}{t^{\beta/\alpha}}\r) 
N_{\alpha}^{\theta}\l(\frac{x}{\tau}\r) 
\frac{d\tau}{\tau} \,,
\, 0 < \beta/\alpha \le 1 \,, \!\!\!\!\!\!
\label{lumapa20}
\ee 
\end{subequations}
or, analogously
\cite{mainardi_etal-fcaa-2001,mainardi_etal-fcaa-2003}: 
\begin{subequations}
\be
\Kabt(x;t) = \int_0^{\infty} \Lat(x;\tau) \, M_\beta(\tau;t) \,
d\tau \,, \quad 0<\beta \le1 \,,
\label{subordinationstfde}
\ee 
\be 
\Kabt(x;t) = \int_0^{\infty} N_{\alpha}^{\theta}(x;\tau) \,
M_{\beta/\alpha}(\tau;t) \, d\tau \,, \quad 0<\beta/\alpha \le 1
\,.
\label{subordinationstfde2}
\ee 
\end{subequations}

The symmetric case, i.e. $\theta=0$, of formula
(\ref{subordinationstfde}) was previosuly derived by Saichev \&
Zavlasky \cite{saichev_etal-c-1997}.  Moreover, combining (\ref{m3})
and (\ref{subordinationstfde}) it follows the identity
\cite{mainardi_etal-fcaa-2001} \be L_{\beta}^{-\beta}(t;\tau) \,
\frac{t}{\beta \, \tau} = M_\beta(\tau;t) \,, \quad 0 < \beta \le 1
\,, \quad \tau \ge 0 \,, \quad t \ge 0 \,, \ee 
and by using self-smilarity properties 
\be 
\hspace{-0.5truecm}
L_{\beta}^{-\beta}\left(\frac{t}{\tau^{1/\beta}}\right) \,
\frac{t}{\beta \, \tau^{1/\beta+1}} = \frac{1}{t^\beta} \,
M_\beta\left(\frac{\tau}{t^\beta}\right) \,, \quad 0 < \beta \le 1 \,,
\,\,\, \tau \ge 0 \,, \,\,\, t \ge 0 \,. \hspace{-0.4truecm}
\label{LMformula}
\ee

\smallskip
Since the relationship among PDFs $\Kabt$, $\Lat$, $M_\nu$ and
$N_\alpha^\theta$ (see Appendix), integral representation formulae
(\ref{subordinationstfde}) and (\ref{subordinationstfde2}) can be
re-stated also in terms of only the particle PDF $\Kabt(x;t)$ with the
opportune choice of parameters
\cite{mainardi_etal-fcaa-2001,mainardi_etal-fcaa-2003}, i.e.
\begin{subequations}
\be \Kabt(x;t) = 2 \, \int_0^{\infty}
K_{\alpha,1}^\theta(x;\tau) \, K_{2,2\beta}^0(\tau;t) \, d\tau \,,
\quad 0<\beta \le1 \,,
\label{subordinationstfde3}
\ee 
\be \Kabt(x;t) = 2 \, \int_0^{\infty}
K_{\alpha,\alpha}^\theta(x;\tau) \, K_{2,2\beta/\alpha}^0(\tau;t) \,
d\tau \,, \quad 0<\beta/\alpha \le 1 \,.  
\ee 
\end{subequations}

Formula (\ref{subordinationstfde}), or the analog ones, shows that the
solution of the space-time fractional diffusion equation
(\ref{STFDE0}) can be expressed in terms of the solution of the
space-fractional diffusion equation of order $\alpha$,
i.e. $K_{\alpha,1}^\theta(x;t)=\Lat(x;t)$, and of the solution of the
time-fractional diffusion equation of order $2\beta$,
i.e. $K_{2,2\beta}^0(\tau;t)=M_\beta(\tau;t)/2$, $\tau \ge 0$.
Moreover, formulae (\ref{subordinationstfde}) and
(\ref{subordinationstfde2}), or the analog ones, by involving
non-negative functions allow for interpreting $\Kabt(x;t)$ as a PDF.
Furthermore, it is worth noting to remark that formula
(\ref{subordinationstfde2}), or the analog ones, is fundamental to
extend such probability interpretation to the range $1 < \beta \le
\alpha \le 2$.

\subsection{A new integral representation formula}
\label{new_integral_formula}

By using formula (\ref{subordinationstfde}) a new integral representation formula for
the space-time fractional diffusion can be derived. 
A preliminary derivation was presented in the proceedings papers \cite{pagnini-pmam-2014,pagnini-mesa-2014}. 

\begin{theorem}\label{Th1}
Let $\Kabt(x;t)$ be the fundamental solution 
of the space-time fractional diffusion equation (\ref{STFDE0}) with
initial and boundary conditions $u(x;0)=\delta(x)$ and $u(\pm\infty;t)=0$
and parameters $\alpha$, $\theta$, $\beta$ such that
$$
0 < \alpha \le 2 \,, \quad |\theta| \le \min\{\alpha,2-\alpha\} \,, \quad 0 < \beta \le 1 \,,
$$ 
then the following integral representation formula holds true for $0 < x < \infty$:
\be 
\hspace{-0.1truecm}
\Kabt(x;t)= \int_0^\infty
L_{\eta}^{\gamma}(x;\xi) \, K_{\nu,\beta}^{-\nu}(\xi,t) \, d\xi \,,
\quad {\rm with} \quad \alpha=\eta \nu \,, \quad \theta=\gamma \nu \,,
\label{subordinationnew}
\ee 
and 
$$
0 < \eta \le 2 \,, \quad |\gamma| \le \min\{\eta,2-\eta\} \,, 0 < \nu \le 1 \,.
$$
\end{theorem}
\proof
It is known that the following integral representation formula for 
L\'evy stable density holds
\cite{feller-1971,mainardi_etal-fcaa-2003,mainardi_etal-jms-2006}: 
\be
\Lat(x;t)=\int_0^\infty L_{\eta}^{\gamma}(x;\xi) L_{\nu}^{-\nu}(\xi;t)
\, d\xi \,, \quad \alpha=\eta \nu \,, \quad \theta=\gamma \nu \,,
\label{subordinationlevy}
\ee 
where 
\begin{subequations} 
\be 0 < \alpha \le 2 \,, \quad |\theta| \le
\min\{\alpha,2-\alpha\} \,,
\label{param1}
\ee 
\be 
0 < \eta \le 2 \,, \quad |\gamma| \le \min\{\eta,2-\eta\} \,,
\quad 0 < \nu \le 1 \,.
\label{param2}
\ee 
\end{subequations} 
Hence, inserting (\ref{subordinationlevy}) into formula (\ref{subordinationstfde}) gives:
\begin{eqnarray}
\Kabt(x;t) &=&\int_0^\infty \left\{ \int_0^\infty
L_{\eta}^{\gamma}(x;\xi) L_{\nu}^{-\nu}(\xi;\tau) \, d\xi \right\}
\, \Mb(\tau;t) \, d\tau \,, \nonumber \\ &=&\int_0^\infty
L_{\eta}^{\gamma}(x,\xi) \left\{ \int_0^\infty
L_{\nu}^{-\nu}(\xi;\tau) \Mb(\tau;t) \, d\tau \right\} \, d\xi
\,,
\label{directing}
\end{eqnarray}
where the exchange of integration is allowed by the fact that the
involved functions are normalized PDFs.  
Using again (\ref{subordinationstfde}) to compute the integral into braces in (\ref{directing}), 
since $0 < \beta \le 1$, integral representation formula (\ref{subordinationnew}) is obtained.
\proofend

\begin{corollary}
In the particular case $\eta=2$ and $\gamma=0$, so that $\nu=\alpha/2$ and $\theta=0$, 
the spatial variable $x$ emerges to be distributed according to a Gaussian PDF and
integral representation formula (\ref{subordinationnew}) becomes
\be
\hspace{-0.1truecm}
\Kab(x;t)= \int_0^\infty \mG(x;\xi) \,
K_{\alpha/2,\beta}^{-\alpha/2}(\xi;t) \, d\xi \,, \quad 0 < \alpha
\le 2 \,, \quad 0 < \beta \le 1 \,.
\label{subordination}
\ee
\end{corollary}
%
\proof
The proof straighforwardly follows from the fact that identities in (\ref{gaussiancase}) hold
\be
L_2^0(x;t)=\mG(x;t)=\frac{\e^{-x^2/(4 \, t)}}{\sqrt{4 \, \pi \, t}} \,.
\ee 
\endproof

Formula (\ref{subordination}) involves the Gaussian PDF $G(x,t)$ and can be stochastically 
interpreted in different ways, including that of a subordination process. 
However, in the next section we will follow an alternative interpretation,
which allows for deriving a new self-similar Gaussian-based stochastic solution of the 
symmetric space-time fractional diffusion equation (\ref{STFDE}) with Gaussian stationary increments.

\section{The stochastic solution with Gaussian stationary increments of the 
symmetric space-time fractional diffusion equation}
\label{sec:hsssi}
\setcounter{section}{3}
\setcounter{equation}{0}\setcounter{theorem}{0}

In this Section we propose a novel approach to obtain a new stochastic
solution $\Xab(t)$ of the symmetric space-time fractional diffusion
equation (\ref{STFDE}) which
has the valuable property to have stationary increments and to be fully characterized.
This method is based on the correspondence of the PDFs
resulting from the integral representation formula (\ref{subordination}) 
with the PDFs resulting from the product of two independent random variables. 

This approach was preliminarly presented in the proceedings paper \cite{pagnini-mesa-2014}.

\smallskip
In general, the one-point one-time PDF is not sufficient to characterise a stochastic process.
In fact, starting from a Cauchy problem whose solution can be interpreted as a PDF,
there is an infinity of stochastic processes that share the same one-dimensional distribution.
But, in the present approach, 
the correspondence with the product of an appropriate random variable and a Gaussian stochastic process,
suggests a method to solve this indeterminacy.
In particular, because the Gaussian process is fully charaterised. 

\smallskip
It is well known that the PDF of the product of two independent random 
variables is given by an integral formula
\cite{feller-1971,mainardi_etal-fcaa-2003,mainardi_etal-jms-2006}.
The correspondence between a parent-directing subordination and the 
product of two independent random variables for self-similar processes was 
also highlighted in Reference \cite{pagnini_etal-ptrsa-2013}.  
This relationship is useful to establish classes of 
{\it Hurst Self-Similar with Stationary Increments} (H-SSSI)
stochastic processes to model anomalous diffusion \cite{pagnini_etal-ptrsa-2013}.
In this respect the following definition is reminded:
\begin{definition}[H-SSSI processes]
A stochastic process $W(t)$, $t \ge 0$, with values in $R$, is a 
H-SSSI process if:
(i) it is a self-similar process, i.e.: $W(at)$ and $a^H W(t)$
have the same finite-dimensional distributions for all $a > 0$, 
where $H$ is the Hurst exponent;
%
%
(ii) it has stationary increments, i.e., the distribution of the increments 
$W(t+\tau)-W(t)$ is invariant under the time shift transformation:
$t \rightarrow t+s$.
%
\end{definition}

First, the following lemma is given.
\begin{lemma}
Let $Z_1$ and $Z_2$ be two real independent random variables
whose PDFs are $p_1(z_1)$ and $p_2(z_2)$, respectively, with $z_1 \in
R$ and $z_2 \in R^+$. 
Let $Z$ be the random variable obtained by the product of
$Z_1$ and $Z_2^\gamma$, i.e.:  
\be 
Z = Z_1 \, Z_2^\gamma \,.
\label{Z}
\ee 
Then, denoting with p(z) the PDF of $Z$, it results:
\be 
p(z)=\int_0^\infty
p_1\left(\frac{z}{\lambda^\gamma}\right) p_2(\lambda) \,
\frac{d\lambda}{\lambda^\gamma} \,.
\label{pdfproduct}
\ee 
\end{lemma}
%
%
\proof
The joint PDF of $Z_1$ and $Z_2$ is $p(z_1, z_2)=p_1(z_1)p_2(z_2)$ and their joint probability 
to be in the intervals $z_1 < Z_1 < z_1 + dz_1$ and $z_2 < Z_2 < z_2 + dz_2$
is given by $p(z_1, z_2) dz_1 dz_2$, where $dz_1 dz_2$ is an infinitesimally small area.
Relationship (\ref{Z}) suggests the following variable transformation:
\be
z_1=z/\lambda^\gamma\, ; \quad z_2=\lambda \,.
\nonumber
\ee
After the substitution of the above change of variables into the formula of the probability, 
it follows:
\be
p(z_1,z_2) \, d z_1 dz_2 = p_1(z/\lambda^\gamma)
p_2(\lambda) \, J \, d z d\lambda\, ,
\nonumber
\ee
%
%
where $J=1/\lambda^\gamma$ is the Jacobian of the transformation.
Finally, integration in $d\lambda$ gives formula (\ref{pdfproduct}).
\proofend

\smallskip
It is here reminded that there is a formal correspondence between the PDF of the product of two
variables as stated in formula (\ref{pdfproduct}) and the integral representation formula 
(\ref{subordination}), 
such that it is possible to link the mathematical expressions of the two PDFs. 

\smallskip
By applying the changes of variables $z=x t^{-\gamma \omega}$ and $\lambda=\tau t^{-\omega}$, 
integral representation (\ref{subordination0}) is recovered from (\ref{pdfproduct}) by setting
\be 
\hspace{-0.2truecm}
\frac{1}{t^{\gamma \omega}} p \left(\frac{x}{t^{\gamma \omega}}\right) 
\equiv p(x;t) \,, \quad \!\! 
\frac{1}{\tau^\gamma} p_1
\left(\frac{x}{\tau^\gamma}\right) \equiv \psi(x;\tau) \,, \quad \!\!
\frac{1}{t^\omega} p_2 \left(\frac{\tau}{t^\omega}\right) \equiv
\varphi(\tau;t) \,.
\label{equality}
\ee
Then, identifying functions and parameters in (\ref{equality}) as follows
\begin{subequations} 
\be p(z) \equiv \Kab(z) \,, \quad p_1(z_1)
\equiv \mG(z_1) \,, \quad p_2(z_2) \equiv
K_{\alpha/2,\beta}^{-\alpha/2}(z_2) \,,
\label{equiv}
\ee 
\be 
\gamma=1/2 \,, \quad \omega=2\beta/\alpha \,, \quad
\gamma\omega=\beta/\alpha \, , 
\ee 
\end{subequations} 
formula (\ref{pdfproduct}) reduces to the new integral representation formula
(\ref{subordination}) for the symmetric space-time fractional diffusion
equation.

The correspondence between the two above mechanisms, 
namely formulae (\ref{subordination0}) and (\ref{pdfproduct}),
can be understood as follows \cite{pagnini_etal-ptrsa-2013}.
Let $W(t)$, $t \ge 0$, a H-SSSI process.
Hence, if parameter $a$ is turned into a random
variable, in the parent-directing subordination approach, the
resulting process emerges to be $X(t)=Y(T(t))=W(at)$ where it holds
$Y(\tau)=W(\tau)$ and $\tau=T(t)=a \, t$, and, in the approach based
on the product of two independent random variables, the resulting
process is $Z(t)=Z_2^\gamma \, Z_1(t)=a^H W(t)$ where it holds
$Z_2^\gamma=a^H$ and $Z_1(t)=W(t)$.  Due to the self-similarity nature
of $W(t)$, processes $X(t)=W(at)$ and $Z(t)=a^HW(t)$ have the same
finite-dimensional distributions.  This means that the process $Z(t)$
has the same single-point single-time density of a subordinated stochastic process
where the parent process $Y(\tau)$ is a self-similar process,
i.e. $Y(\tau)=W(\tau)$, and the operational time $\tau$ is a line with
stochastic slope, i.e. $\tau=T(t)=a \, t$.  

\smallskip
In terms of random variables it follows that
\be
Z=X \, t^{-\beta/\alpha} \q {\rm and} \q Z=Z_1 Z_2^{1/2} \,, 
\ee
hence it holds 
\be
X=Z \, t^{\beta/\alpha}=Z_1 \, t^{\beta/\alpha} \, Z_2^{1/2}
=G_{2\beta/\alpha}(t) \, \sqrt{\Lambda_{\alpha/2,\beta}} \,.  
\ee
Since $p_1(z_1) \equiv \mG(z_1)$, $Z_1$ is a Gaussian random variable (see 
(\ref{equiv})). Consequently, the stochastic process $Z_1
t^{\beta/\alpha}=G_{2\beta/\alpha}(t)$ is a Gaussian process displaying anomalous
diffusion. 
Further, the random variable $Z_2=\Lambda_{\alpha/2,\beta}$ is distributed 
according to $p_2(z_2) \equiv K_{\alpha/2,\beta}^{-\alpha/2}(z_2)$ (see (\ref{equiv})).

\smallskip
The above reasoning is based on the same constructive approach adopted by Mura 
\cite{mura-phd-2008} 
to built up the {\it generalized grey Brownian motion}
\cite{mura-phd-2008,mura_etal-jpa-2008,mura_etal-itsf-2009,dasilva_etal-s-2014,dasilva-ijmpcs-2015}. 
In summary, we define the following class of H-SSSI processes:
\begin{definition}[Gaussian-based H-SSSI stochastic processes]
Let $X_{\alpha,\beta}(t)$, $t\ge 0$, be a H-SSSI process defined
by 
\be \Xab(t)= \sqrt{\Lambda_{\alpha/2,\beta}} \,\,
G_{2\beta/\alpha}(t) \,, 
\quad 0 < \beta \le 1 \,,
\quad 0 < \alpha \le 2 \,, 
\label{repstBm}
\ee 
where the stochastic process $G_{2\beta/\alpha}(t)$ is a H-SSSI Gaussian
process with power law variance $t^{2\beta/\alpha}$ 
and $\Lambda_{\alpha/2,\beta}$ 
is an independent constant non-negative random variable distributed according to
the PDF $K_{\alpha/2,\beta}^{-\alpha/2}(\lambda)$, $\lambda \ge 0$,  
that is a special case of (\ref{subordinationstfde}).
Then we say that $\Xab(t)$ is a Gaussian-based H-SSSI stochastic process.
\end{definition}

The following theorem can be now proved.
\begin{theorem}[Stochastic solution of equation (\ref{STFDE})]
\label{main_result}
The parametric class of Gaussian-based H-SSSI stochastic processes $\Xab(t)$ defined in (\ref{repstBm}), 
and depending on the paramenters $0 < \beta \le 1$ and $0 < \alpha \le 2$,
is a class of stochastic solutions of the symmetric space-time fractional diffusion equation
(\ref{STFDE}).
This means that the one-time one-point PDF of $\Xab(t)$ is the fundamental solution of
equation (\ref{STFDE}), namely the PDF $K_{\alpha,\beta}^0(x;t)$ defined in (\ref{subordination}).
\end{theorem}
\proof
For the given sequence of times $(t_1, ... , t_n)$, 
the joint PDF $f_{\alpha,\beta}$ of the $n$-dimensional particle
random vector ${\bf X} = (\Xab(t_1), ... ,\Xab(t_n))$ to be in 
the position vector ${\bf x}=(x_1,x_2,...,x_n)$ 
can be derived from (\ref{pdfproduct}) and (\ref{equiv})
by using the Kolmogorov extension theorem and it results to be
\begin{eqnarray} 
f_{\alpha,\beta}(x_1,x_2,\dots,x_n; \gamma_{\alpha,\beta})= \hspace{7.0truecm} \nonumber \\
\displaystyle
\frac{1}{\sqrt{(2 \pi \lambda)^n \det{\gamma_{\alpha,\beta}}}}
\int_{0}^{\infty}
\exp\left\{-\frac{1}{2 \lambda} \bx^T [\gamma_{\alpha,\beta}]^{-1} \bx\right\}
K_{\alpha/2,\beta}^{-\alpha/2}(\lambda) \, d\lambda \,,
\label{stBm}
\end{eqnarray} 
where $\gamma_{\alpha,\beta}(t_i,t_j)$ is the covariance matrix of the random vector
corresponding to the Gaussian process $G_{2\beta/\alpha}(t)$ and ${[\gamma_{\alpha,\beta}]}^{-1}$ is the
inverse covariance matrix.
The one-time one-point PDF ($n=1$) is
\be 
f_{\alpha,\beta}(x;t) =
\int_{0}^{\infty} \frac{1}{\lambda^{1/2}} \mG\left(\frac{x \,
  t^{-\beta/\alpha}}{\lambda^{1/2}}\right) \,
K_{\alpha/2,\beta}^{-\alpha/2}(\lambda) \, d\lambda = \Kab(x \,
t^{-\beta/\alpha}) \,,
\label{superpositionlambda}
\ee 
where the scaling relationship $x/t^{\beta/\alpha}$, which drives the anomalous diffusion scaling, 
has been taken into account.

\smallskip
Formula (\ref{superpositionlambda}) is the mathematical expression of the 
one-time one-point PDF of the Gaussian-based H-SSSI stochastic process 
$\Xab(t)$ given in (\ref{repstBm}).
After the change of variable $\lambda=\tau \, t^{-2\beta/\alpha}$, 
formula (\ref{superpositionlambda}) can be rewritten in the following way:
\be 
\int_{0}^{\infty} \frac{1}{\tau^{1/2}} \,
\mG\left(\frac{x}{\tau^{1/2}}\right) \,
K_{\alpha/2,\beta}^{-\alpha/2}\left(\frac{\tau}{t^{2 
\beta/\alpha}}\right) \, \frac{d\tau}{t^{2\beta/\alpha}} =
t^{-\beta/\alpha}\Kab\left(\frac{x}{t^{\beta/\alpha}}\right) \,. 
\ee
Considering self-similarity scaling,
this expression is the same given in formula (\ref{subordination}), 
which is the integral representation formula for the solution of the symmetric space-time
fractional diffusion equation (\ref{STFDE}).

Then, the equivalence of the two expressions (\ref{superpositionlambda}) and 
(\ref{subordination}) demonstrates that the H-SSSI Gaussian-based
stochastic process $\Xab(t)$, defined in equation (\ref{repstBm}), is a stochastic solution
of the symmetric space-time fractional diffusion equation (\ref{STFDE}),
thus proving Theorem \ref{main_result}.
\endproof

\smallskip
It is worth noting that formula (\ref{superpositionlambda}) represents a 
superposition of Gaussian processes depending on the value of the 
random multiplicative factor $\sqrt{\Lambda_{\alpha/2,\beta}}$
and that the process $\Xab(t)$ is Gaussian conditional on the value of this multiplicative factor.
The multiplicative factor can be understood, for example, as related to the diffusion
coefficient in analogy with ideas discussed elsewhere
\cite{pagnini_etal-ijsa-2012,pagnini-pa-2014}.

\smallskip
The stochastic process $\Xab(t)$ stated in (\ref{repstBm}) generalizes Gaussian 
processes, which are recovered when $\alpha=2$ and $\beta=1$.  Similarly to
Gaussian process, even this process is uniquely determined by the mean
and the autocovariance structure.
This property directly follows from the fact that
$G_{2\beta/\alpha}(t)$ is a Gaussian stochastic process and 
$\Lambda_{\alpha/2,\beta}$ is an independent constant non-negative random variable.  

\smallskip
Before to discuss the numerical simulations 
it is highlighted that any Gaussian process $G_{2\beta/\alpha}(t)$
can be used to build up the stochastic solution (\ref{repstBm}),
because each Gaussian process is characterized by a particular covariance matrix.
The following numerical simulations are performed by choosing 
the standard fBm with Hurst exponent $H=\beta/\alpha < 1$ as Gaussian stochastic process $G_{2\beta/\alpha}(t)$. 
This choice constraints parameters $\alpha$ and $\beta$ to fall inside the intervals
\be
0 < \beta \le 1 \,,
\quad 0 < \beta < \alpha \le 2 \,.
\ee
Note that after this choice the stochastic trajectories $\Xab(t)$ 
have Gaussian stationary increments.

\section{Numerical simulations}
\label{sec:simulations}
\setcounter{section}{4}
\setcounter{equation}{0}
\setcounter{theorem}{0}
\setcounter{figure}{0}
The fBm is the natural choice of the Gaussian process $G_{2\beta/\alpha}(t)$
in the formulation of the the Gaussian-based H-SSSI stochastic process $\Xab(t)$ 
defined in (\ref{repstBm}).
The foundation of $\Xab(t)$ on the fBm
is a remarkable property, as it allows to
reduce a large number of issues to the analysis of the fBm, which has
been largely studied (see, e.g., Reference \cite{biagini_etal-2008}).  
Moreover, the stationarity of increments makes the stochastic process
(\ref{repstBm}) more efficient for the simulations of stochastic
trajectories by means of Monte Carlo methods.

\smallskip
Now, given a set of time points $(t_1, t_2, ..., t_n)$, the fBm is defined by 
the following covariance matrix:
\be 
\gamma_{\alpha,\beta}(t_i,t_j)=
t_i^{2\beta/\alpha}+t_j^{2\beta/\alpha}-|t_i-t_j|^{2\beta/\alpha} \,, \quad i,j=1,\dots,n \,.  
\ee 
In order to perform numerical simulations of the stochastic solution 
$\Xab(t)$ stated in equation (\ref{repstBm}), both the generation of the random variable
$\Lambda_{\alpha/2,\beta}$, distributed according to
$K_{\alpha/2,\beta}^{-\alpha/2}(\lambda)$, 
and the fBm $G_{2\beta/\alpha}(t)$ are here discussed.

\smallskip
From (\ref{directing}) and (\ref{subordinationnew}), according to
(\ref{subordinationstfde}), it follows: \be
K_{\alpha/2,\beta}^{-\alpha/2}(\xi;t)= \int_0^\infty
L_{\alpha/2}^{-\alpha/2}(\xi;\tau) \, \Mb(\tau;t) \, d\tau \, ,
\quad 0 < \beta \le 1 \,, \ee and, using the self-similarity property: 
\be 
\hspace{-0.6truecm}
t^{-2\beta/\alpha}
K_{\alpha/2,\beta}^{-\alpha/2} \! \left(\frac{\xi}{t^{2\beta/\alpha}}\right)
\!=\!\!
\int_0^\infty \!\!\!
L_{\alpha/2}^{-\alpha/2} \! \left(\frac{\xi}{\tau^{2/\alpha}}\right)
\! \Mb\left(\frac{\tau}{t^\beta}\right) 
\! \frac{d\tau}{\tau^{2/\alpha} \, t^\beta}
\,, 0 < \beta \le 1 \,. \!\!\!\!\!\!\! 
\ee 
After the changes of variable
$\xi=t^{2\beta/\alpha} \lambda$ and $\tau=t^\beta y$, it holds: \be
K_{\alpha/2,\beta}^{-\alpha/2}(\lambda)= \int_0^\infty
L_{\alpha/2}^{-\alpha/2}\left(\frac{\lambda}{y^{2/\alpha}}\right)
\Mb(y) \, \frac{d y}{y^{2/\alpha}} \,, \quad 0 < \beta \le 1 \,.
\label{directing2}
\ee 
The representation integral (\ref{directing2}) suggests that
$\Lambda_{\alpha/2,\beta}$ can be computed by means of the product
of two independent random variables, see Section \ref{sec:hsssi}, i.e.:
\be \Lambda_{\alpha/2,\beta}=\Lambda_1 \cdot \Lambda_2 =
\mLe_{\alpha/2} \cdot \mMb^{2/\alpha} \,,
\label{randomLambda}
\ee where $\Lambda_1=\mLe_{\alpha/2}$ and $\Lambda_2=\mMb$ are
distributed according to the extremal stable density
$L_{\alpha/2}^{-\alpha/2}(\lambda_1)$ and to the density
$\Mb(\lambda_2)$, respectively, so that $\lambda=\lambda_1 \,
\lambda_2$.

\smallskip
Then, from (\ref{LMformula}) and setting $t=1$, 
it follows that the random variable $\mMb$ can be determined by an extremal stable random
variable according to \cite{cahoy-cstm-2012}
\be 
\mMb = \left[\mLe_{\beta}\right]^{-\beta} \,.
\label{randomM}
\ee 
Finally, the random variable $\Lambda_{\alpha/2,\beta}$ is computed by the product 
\be 
\Lambda_{\alpha/2,\beta}= \mLe_{\alpha/2}
\cdot \left[\mLe_{\beta}\right]^{-2\beta/\alpha} \,.
\label{randomLambda2}
\ee 

\smallskip
In summary, the stochastic solution (\ref{repstBm}) of the symmetric
space-time fractional diffusion equation (\ref{STFDE}) is numerically
simulated by the following process 
\be 
X_{\alpha,\beta}(t)=
\sqrt{\mLe_{\alpha/2}} \cdot \left[\mLe_\beta \right]^{-\beta/\alpha}
\cdot G_{2\beta/\alpha}(t) \,, 
\ee 
and the random generation is discussed below.

\smallskip
The computer generation of extremal stable random
variables of order $0 < \mu < 1$ is obtained by using the well-known
method by Chambers, Mallows and Stuck \cite{chambers_etal-jasa-1976,weron-spl-1996} 
\be 
\hspace{-0.4truecm}
\mLe_\mu=
\frac{\sin[\mu(r_1+\pi/2)]}{(\cos r_1)^{1/\mu}}
\left\{\frac{\cos[r_1-\mu(r_1+\pi/2)]}{-\ln r_2}\right\}^{(1-\mu)/\mu} \,,
\,\, 0 < \mu < 1 \,, 
\hspace{-0.3truecm}
\label{randomL}
\ee 
where $r_1$ and $r_2$ are random variables
uniformly distributed in $(-\pi/2,\pi/2)$ and $(0,1)$, respectively.

\smallskip
Regarding the fBm $G_{2H}(t)$, with Hurst exponent $H=\beta/\alpha$
and variance $\langle G_{2H}^2 \rangle=2 \, t^{2H}$, the Hosking
direct method is here applied for the range $0<H<1$
\cite{hosking-wrr-1984, dieker-phd-2004}.  In particular, the
so-called fractional Gaussian noise (fGn) is firstly generated following 
its definition over the set of integer numbers:
\be 
Y_{2H}(n)=G_{2H}(n+1)-G_{2H}(n) \Leftrightarrow
G_{2H}(n+1)=G_{2H}(n)+Y_{2H}(n) \,.
\label{fGn}
\ee 

Finally the fBm is generated as a sum of
stationary increments, which are generated according to the following
stationary auto-correlation function, defined over integer numbers
($n\geq 0$): 
\begin{eqnarray} 
\Gamma_{x,2H}(n) 
&=& \langle Y_{2H}(k) Y_{2H}(k+n) \rangle \nonumber \\
&=& \frac12 \left[|n-1|^{2H}-|n|^{2H}+|n+1|^{2H} \right] \,.
\label{fGn_corr}
\end{eqnarray} 

\smallskip
By implementing the Hosking method, a set of stochastic
trajectories, i.e., sample paths, according to the fBm with Hurst
exponent $H=\beta/\alpha$ were generated.  In order to obtain the
corresponding trajectory of the stochastic process of
equation (\ref{repstBm}), each fBm sample path was then multiplied by the
multiplicative factor $\sqrt{\Lambda_{\alpha/2,\beta}}$ .  It is worth
noting that $\Lambda_{\alpha/2,\beta}$ is not a stochastic process
evolving in time, but a constant random variable characterizing the 
random wideness scale of the single sample path.  

\smallskip
The numerical simulations were carried
out, following the Monte Carlo approach, by means of pseudo-random
generators.  For a given set of parameter values ($\alpha$,$\beta$),
the number of trajectories generated were $10^4$ and the number of
time steps $10^3$. Following equation (\ref{fGn}), a unitary time step was
used as a natural choice to generate the stationary increments of fBm.
Changing the time scale requires changing the time step, and the
associated trajectories can be simply derived without any further
numerical simulations by exploiting the self-similar property.
%
%
\begin{figure}[ht]
\begin{center}
\includegraphics[height=0.45\linewidth,width=\linewidth]{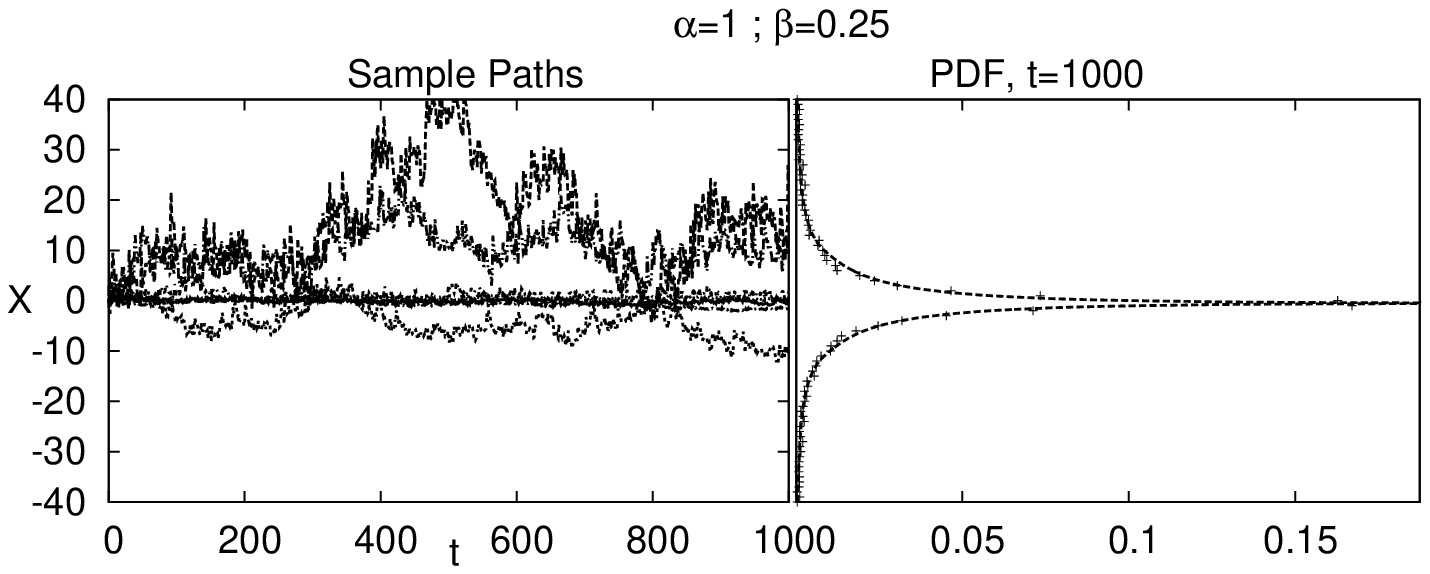}
\includegraphics[height=0.45\linewidth,width=\linewidth]{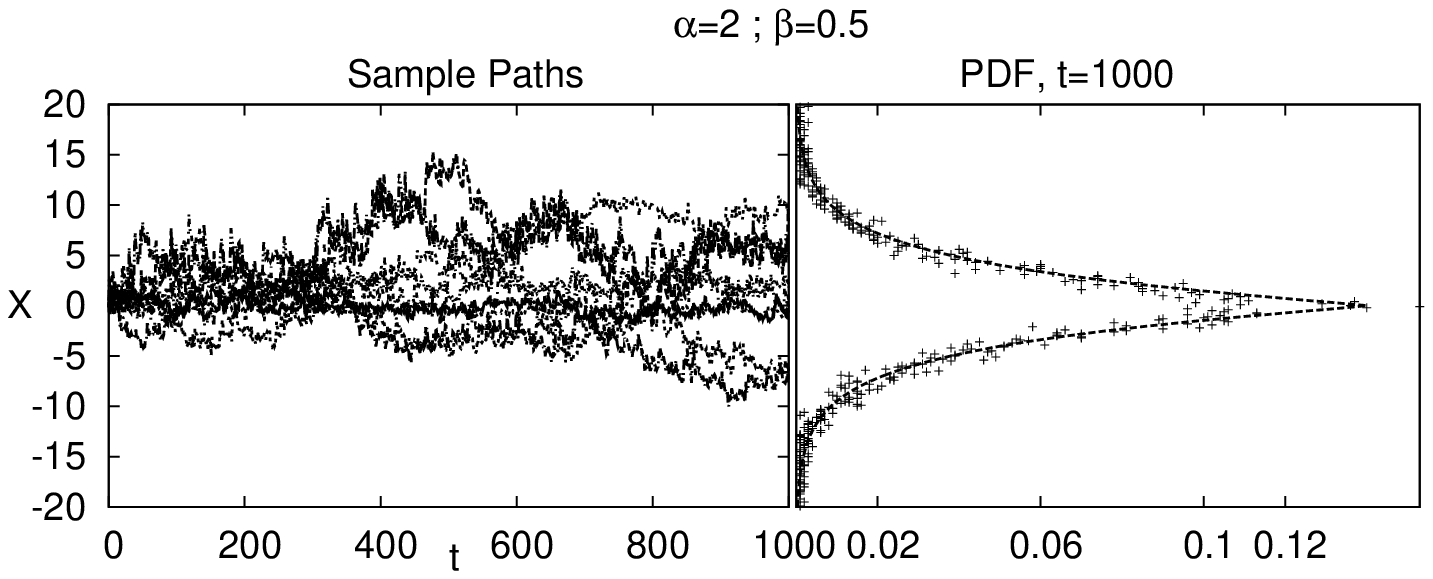}
\caption{Comparison for different couples of parameters $(\alpha,\beta)$ but 
the same $H=\beta/\alpha$,
here $H=\beta/\alpha=0.25$. 
Sample paths are reported on the left panels. In the right panels 
the PDFs of the simulated stochastic processes $X_{\alpha,\beta}(t)$ defined in 
(\ref{repstBm}) are compared with the analytical solutions of
Eq. (\ref{STFDE0}).}
\label{fig1}
\end{center}
\end{figure}
\begin{figure}[ht]
\begin{center}
\includegraphics[height=0.45\linewidth,width=\linewidth]{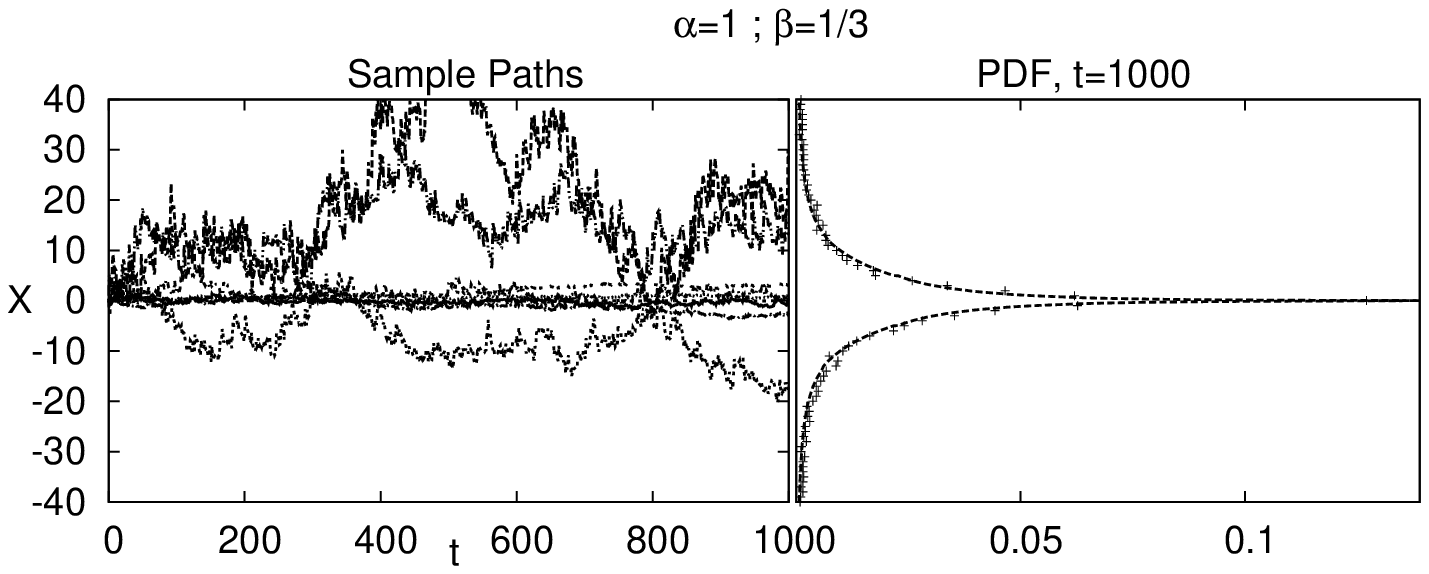}
\includegraphics[height=0.45\linewidth,width=\linewidth]{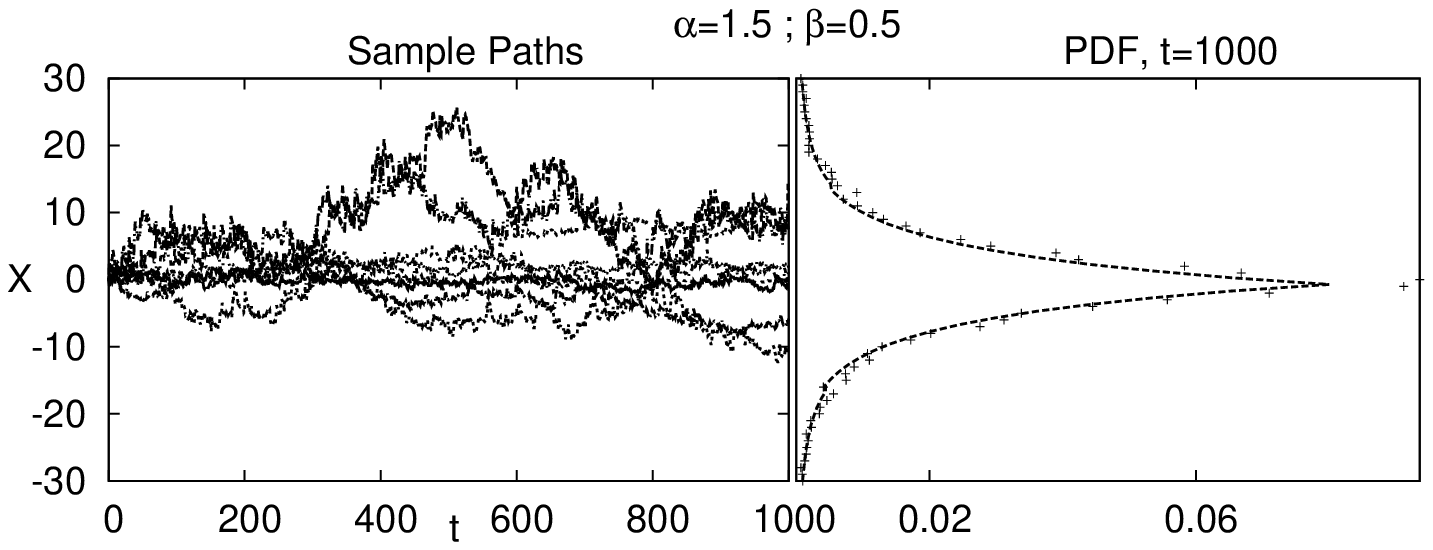}
\caption{Same as Fig. \ref{fig1} with $H=\beta/\alpha=1/3$.}
\label{fig2}
\end{center}
\end{figure}
\begin{figure}[ht]
\begin{center}
\includegraphics[height=0.45\linewidth,width=\linewidth]{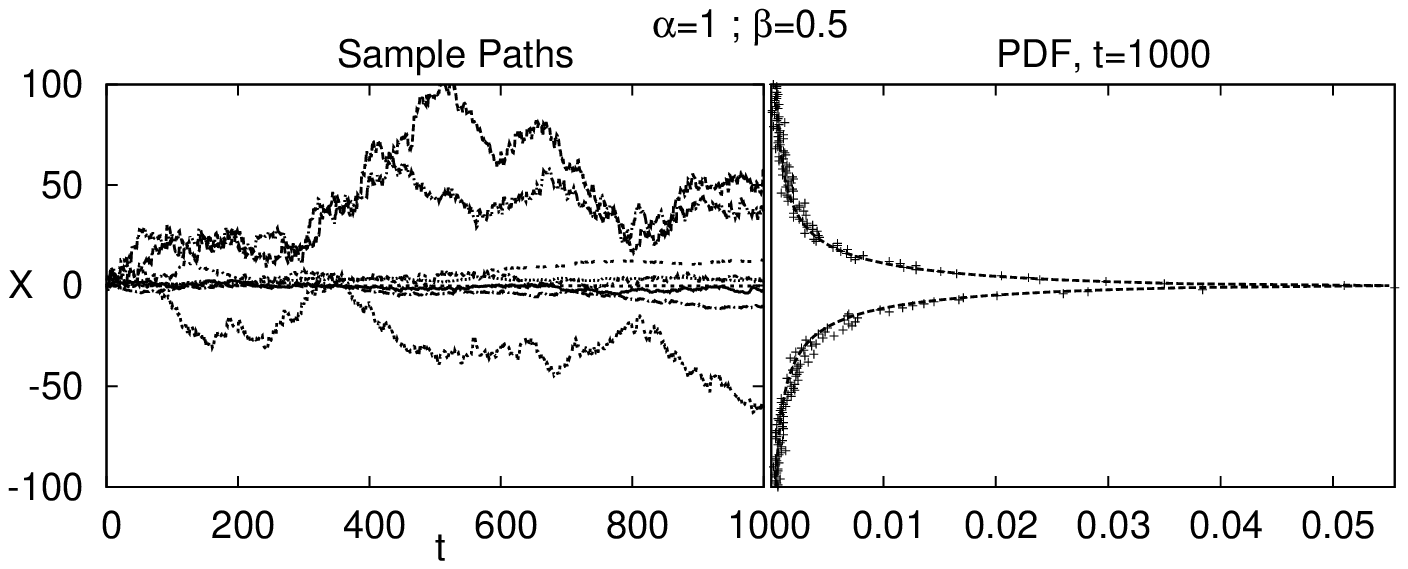}
\includegraphics[height=0.45\linewidth,width=\linewidth]{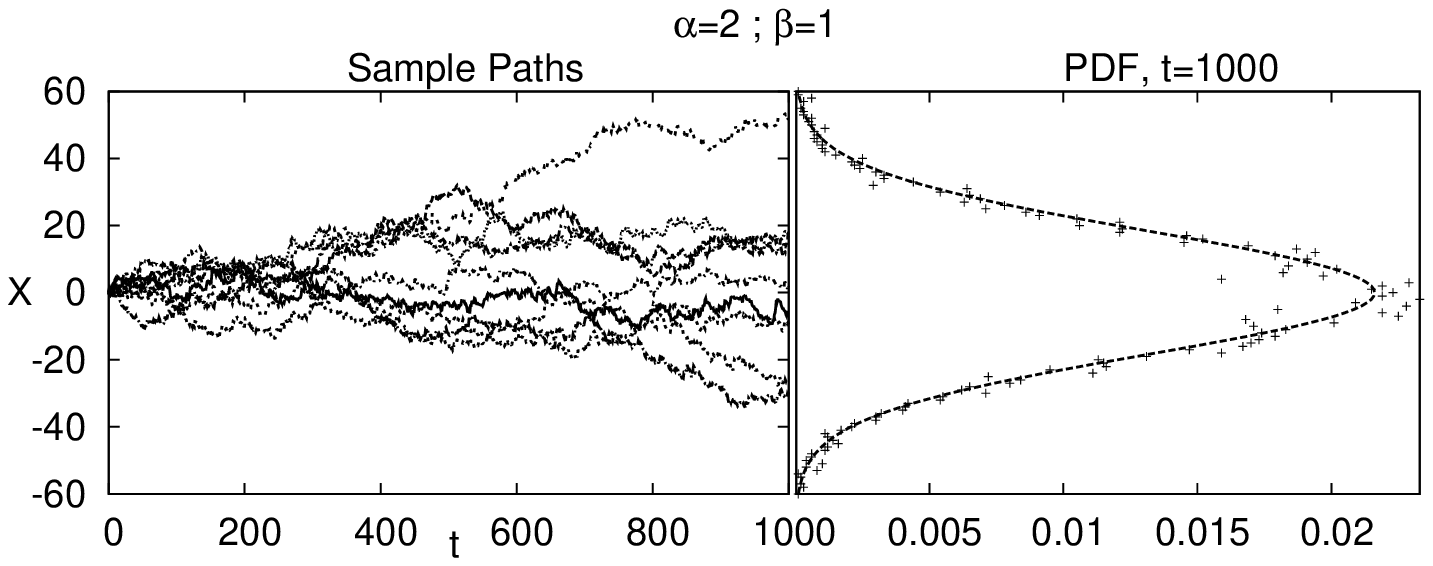}
\caption{Same as Fig. \ref{fig1} with $H=\beta/\alpha=0.5$.}
\label{fig3}
\end{center}
\end{figure}
\begin{figure}[ht]
\begin{center}
\includegraphics[height=0.45\linewidth,width=\linewidth]{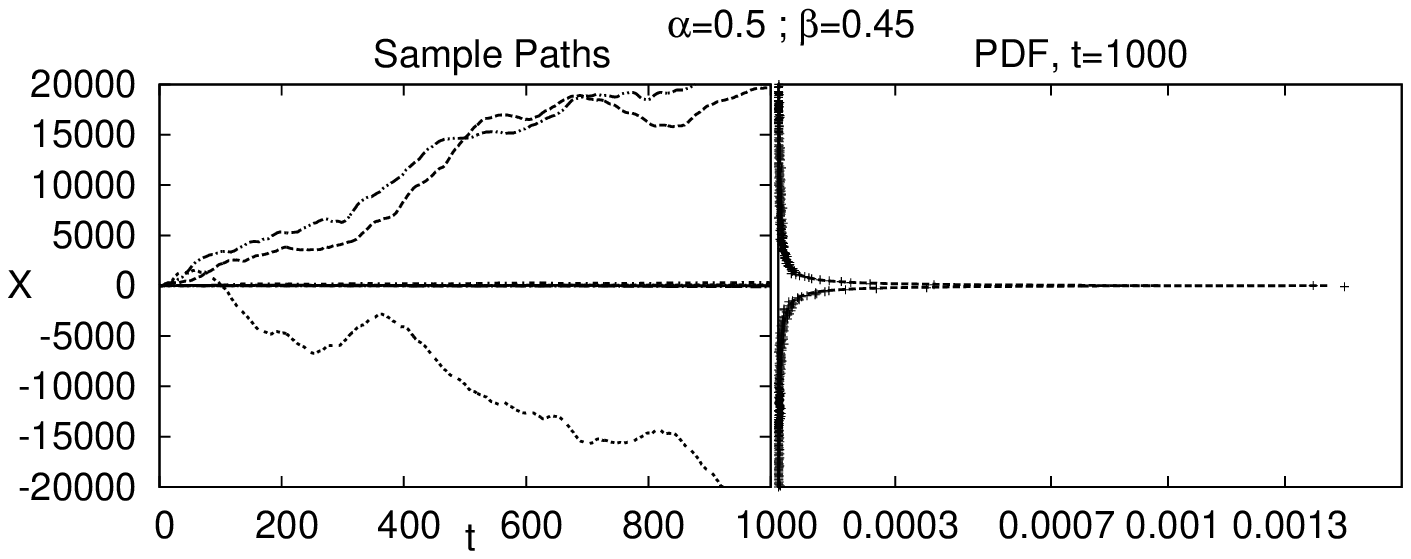}
\includegraphics[height=0.45\linewidth,width=\linewidth]{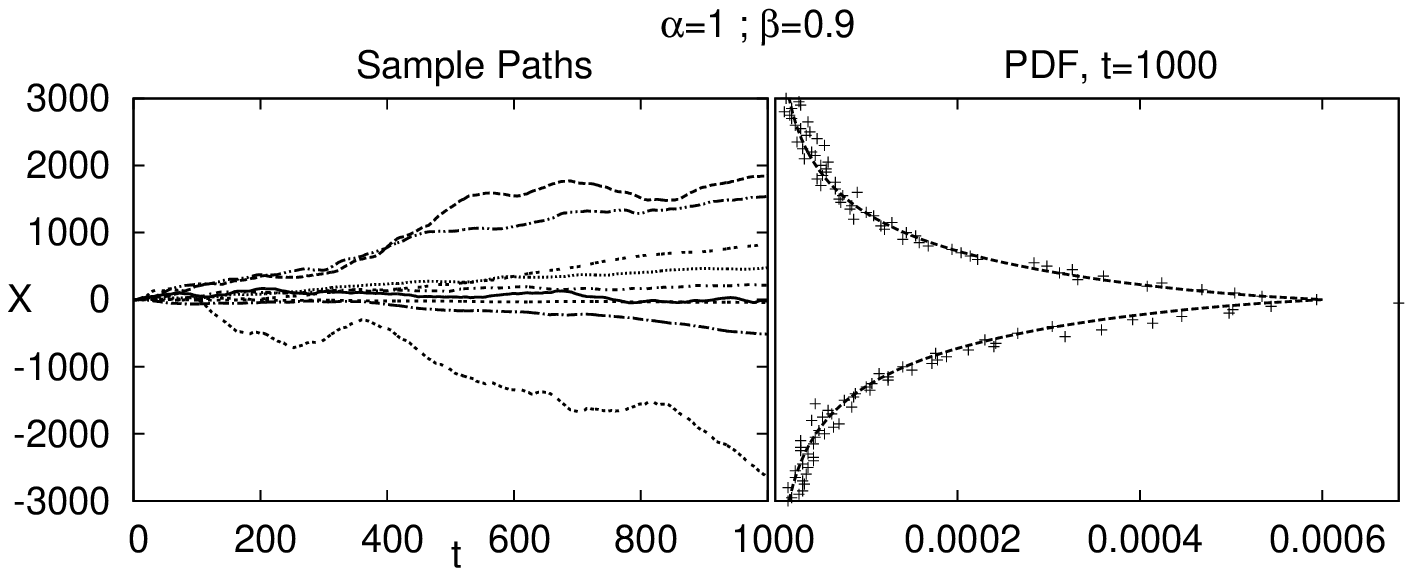}
\caption{Same as Fig. \ref{fig1} with $H=\beta/\alpha=0.9$.}
\label{fig4}
\end{center}
\end{figure}
\begin{figure}[ht]
\begin{center}
\includegraphics[height=0.45\linewidth,width=\linewidth]{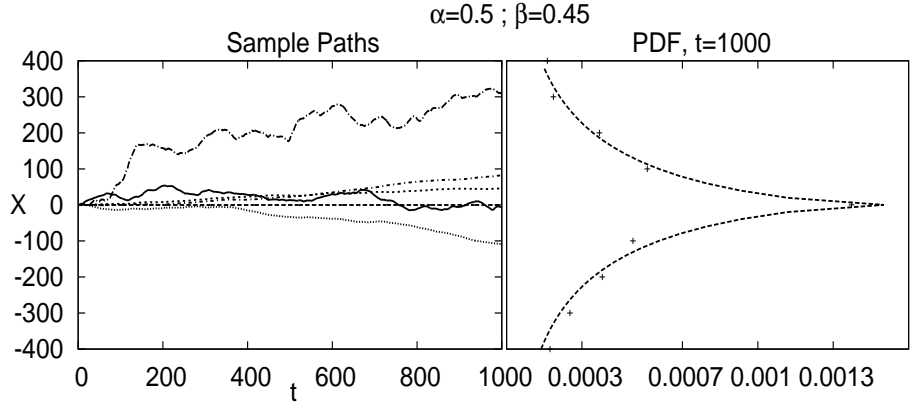}
\caption{$H=\beta/\alpha=0.9$. Zoom of the $X$-range (vertical axis) of the   top 
(both left and right) panels in Fig. \ref{fig4}.}
\label{fig5}
\end{center}
\end{figure}
\begin{figure}[ht]
\begin{center}
\includegraphics[height=0.45\linewidth,width=\linewidth]{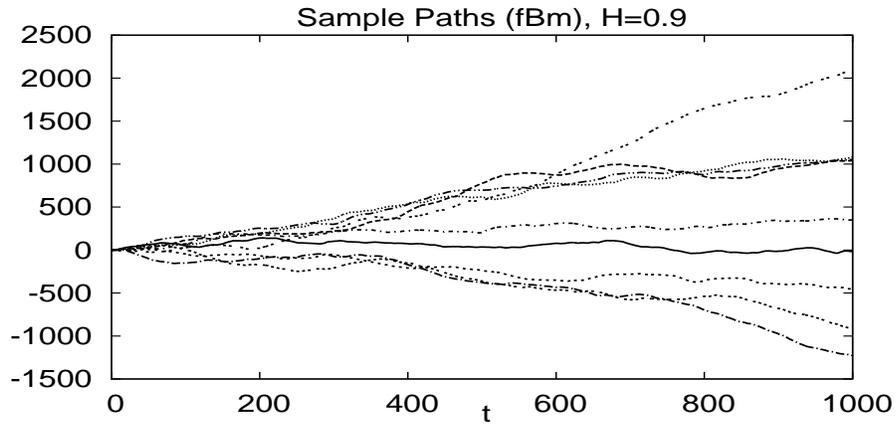}
\caption{Sample paths from the fBm algorithm for $H=0.9$,
to be compared with sample paths of Fig. \ref{fig4} (left panels).}
\label{fig6}
\end{center}
\end{figure}
\begin{figure}[ht]
\begin{center}
\includegraphics[height=0.45\linewidth,width=0.7\linewidth]{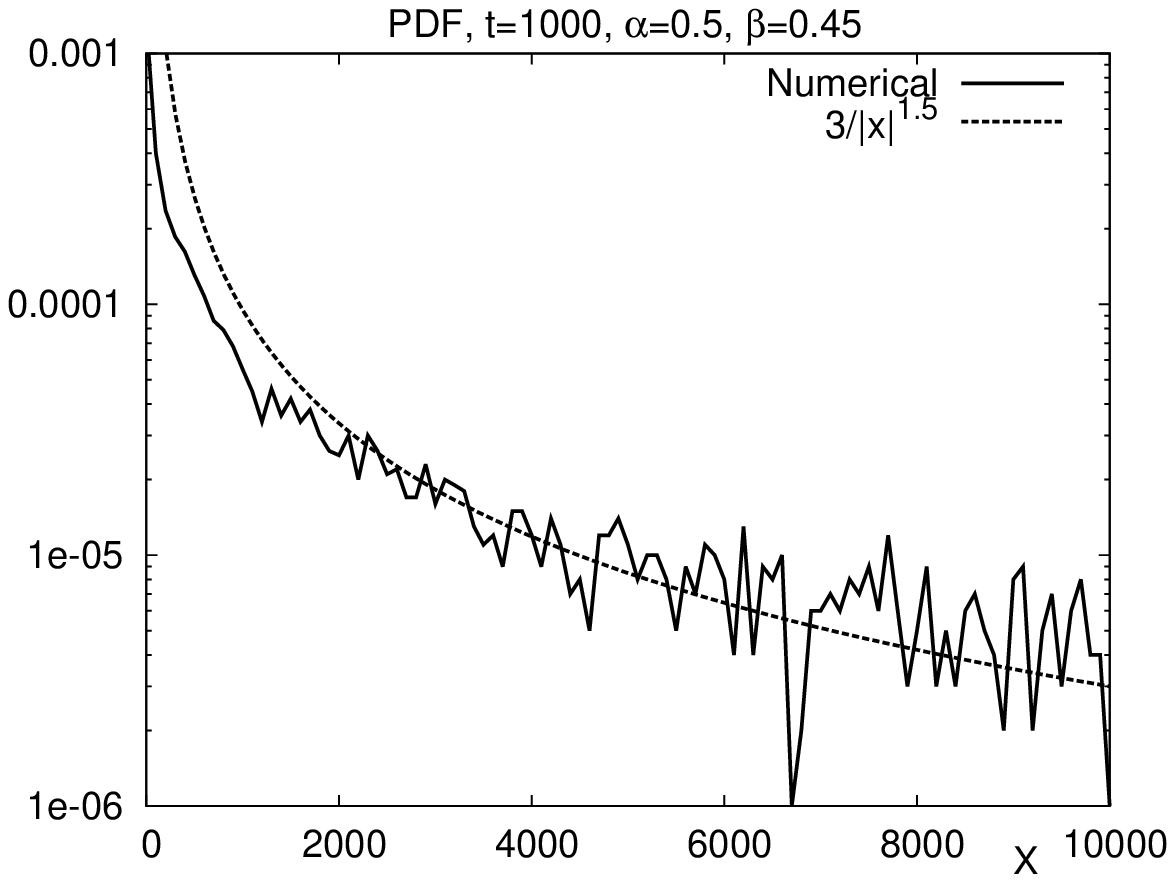}
\caption{$H=\beta/\alpha=0.9, \alpha=0.5$. The PDF evaluated from numerical
  simulations well reproduces the theoretical power-law decay in the
  tail (large $x$): ${\rm PDF}(x)\sim 1/|x|^{\alpha+1}$.}
\label{fig7}
\end{center}
\end{figure}

\smallskip
In Figs. \ref{fig1}-\ref{fig7} the results of numerical simulations have been 
reported for four different values of the Hurst exponent: $H=0.25$, $0.\overline{3}$, $0.5$, $0.9$.  
In Figs. \ref{fig1}-\ref{fig4} two different couples of parameters ($\alpha$,$\beta$),
but giving the same value of $H=\beta/\alpha$, are compared. 
Some sample paths are reported for each case (left panels) together with the 
corresponding PDF (right panels). 
In the right panels the numerical PDFs are compared with the analytical solutions of Eq. (\ref{STFDE0})
by means of a matching between the convergent series
(for $x\rightarrow 0$) and the asymptotic expansions (for $x\rightarrow \pm\infty$)
as derived in \cite{mainardi_etal-fcaa-2001}.
As expected, the spreading of trajectories increases as $H$ increases, with a 
dramatic increase in the neighborhood of $H=1$ (see Fig. \ref{fig4}).
It is interesting to note that, in the process $X_{\alpha,\beta}(t)$, even if 
driven by the fBm $G_{2H}(t)$, the most interesting effects come from the 
random wideness scale $\Lambda_{\alpha/2,\beta}$, which is essentially given by the
multiplication of two extremal L\'evy random variables. This means that a
relatively smooth behaviour in a great portion of sample paths is
observed, in the sense that they show similar random wideness scale going from
one sample path to another. However, a small but non-negligible subset
of sample paths display very different random wideness scales, and this is in fact
related to the random variables $\Lambda_{\alpha/2,\beta}$,
characterizing the ensemble of sample paths.  This can be
seen by comparing Fig. \ref{fig6}, where sample paths from a simple fBm with
$H=0.9$ are reported, with the sample paths in Figs. \ref{fig1}-\ref{fig4}.  
%
In fact,
even if the fBm sample paths show a great spreading (Fig. \ref{fig6}), 
at the same time they show a much higher degree of uniformity of spanning over 
the sample paths with respect to the sample paths shown 
in Figs. \ref{fig1}-\ref{fig4}.
%
%
%
%
%
%

Another fundamental feature of the L\'evy random wideness scale is clarified by 
comparing Fig. \ref{fig4} (top panels) and Fig. \ref{fig5}. Figure \ref{fig5} is a zoom of 
Fig. \ref{fig4} (top panels), where only the sample paths with small wideness scales are reported. 
%
This is caused by the L\'evy-based random wideness scale $\Lambda_{\alpha/2,\beta}$ and it is strongly connected with the emergence of inverse power-law tail
in the range of large $|X|$ (for each fixed time). Notice that the
stochastic solution $X_{\alpha,\beta}(t)$, and the associated
algorithm based on the fBm with stationary increments and a random
wideness scale, well reproduces this power-law decay, as it is shown in 
Fig. \ref{fig7} for the case of $\alpha=0.5$.


%
%

\section{Discussion and conclusions}
\label{sec:conclusion}
In order to provide a new microscopic physical insight to anomalous diffusion,
in the present paper we addressed the problem of finding a Gaussian-based 
stochastic solution of the symmetric space-time fractional diffusion equation 
(\ref{STFDE}).  
The adopted method is based on the fact that the
resulting PDFs from a new integral representation formula 
(\ref{subordinationnew}), which holds for the fundamental solution of
equation (\ref{STFDE0}), and from the product of two independent variables are equal. 
The general integral representation formula 
(\ref{subordinationnew}) turns to be the integral formula (\ref{subordination})
when considering the symmetric space-time fractional diffusion equation
(\ref{STFDE}).
This result is exploited to select a suitable independent constant and non-negative random variable
and a stochastic process that are used to build up,
by means of their product, the Gaussian-based H-SSSI stochastic process (\ref{repstBm}), 
whose one-time one-point PDF is the solution of (\ref{STFDE}).

\smallskip
The fBm has been chosen as a natural prototype for the
Gaussian process driving this stochastic solution, but different choices 
can be made for the Gaussian part. 
Along this line, the
stochastic process (\ref{repstBm}) emerges to be the product of a fBm
and an independent constant and non-negative random variable distributed according to a combination
of two extremal L\'evy distributions.

\smallskip
In literature, several stochastic solutions of the symmetric space-time
fractional equation (\ref{STFDE}) have been derived (see, e.g.,
References \cite{fogedby-pre-1994,meerschaert_etal-pre-2002,scalas_etal-pre-2004,gorenflo_etal-klm-2012}), but, 
up to our knowledge, 
the stochastic process $\Xab(t)$ here presented in (\ref{repstBm}) is the first 
stochastic solution that is also a H-SSSI process based on a Gaussian process with stationary increments.

Moreover, process (\ref{repstBm}) shares with Gaussian processes and
the generalized grey Brownian motion \cite{mura_etal-jpa-2008}, which
is the stochastic solution of the Erd\'elyi--Kober fractional
diffusion equation \cite{pagnini-fcaa-2012,pagnini_etal-ijsa-2012},
the valuable property to be fully characterized by only the first and
second moments of the Gaussian driving jumps 
(i.e., the temporal covariance matrix $\gamma_{\alpha,\beta}(t_i,t_j)$).

\smallskip
The well known characteristic behaviour of anomalous diffusive processes clearly emerge  
from the numerical simulations. 
In particular, when $\alpha \ne 2$, sample paths show very long jumps
and these jumps generate the power-law decay of the PDF.

\smallskip
Finally, valuable spatial and temporal characteristics can be obtained 
with the derived stochastic solution about both the diffusive process and the medium. 
Among these computable characteristics there are, for example, the length and time scales associated to memory
decay and to trapping effects, the diffusion features according to the
Taylor--Green--Kubo formula and also the occurrence of the ergodicity breaking. 
These features cannot be acquired when only the
evolution equation of the particle PDF is known and they will be
investigated with the derived stochastic process in future
developments of the research.

%
%

\appendix

\section*{Appendix: The space-time fractional diffusion equation}
\label{sec:STFDE}
\def\theequation{A.\arabic{equation}}
\setcounter{equation}{0}\setcounter{theorem}{0}

Space-time fractional diffusion equation (\ref{STFDE0}) is obtained
from the ordinary diffusion equation by replacing the first order time
derivative and the second order space drivative with the {\it Caputo}
time-fractional derivative of real order $\beta$ and the {\it Riesz--Feller}
space-fractional derivative of real order $\alpha$, respectively.  The {\it Caputo}
time-fractional derivative $_tD_*^\beta$ is defined by its Laplace
transform as \be \int_0^{+\infty} \e^{-st} \left\{_tD_*^\beta \,
u(x;t) \right\} \, d t = s^\beta \, \widetilde{u}(x;s)
-\sum_{n=0}^{m-1} s^{\beta -1 -n} \, u^{(n)}(x;0^+) \,, \ee with $m-1<
\beta \le m$ and $m \in N$.  The {\it Riesz-Feller} space-fractional
derivative $_xD_\theta^\theta$ is defined by its Fourier trasform
according to \be \int_{-\infty}^{+\infty} \e^{+i \kappa x}
\left\{_xD_\theta^\alpha \, u(x;t) \right\} \, d x = -
|\kappa|^\alpha \, \e^{\displaystyle{i(\sign \kappa)\theta\pi/2}} \,
\widehat{u}(\kappa;t) \,, \ee with $\alpha$ and $\theta$ as in
(\ref{def:alphabetatheta}).

\smallskip
In literature the time-fractional derivative is sometimes considered in the Riemann--Liouville sense. 
The relationship of the time-fractional Riemann--Liouville derivative with the
time-fractional derivative in the Caputo sense is the following
\cite{gorenflo_etal-klm-2012} \be {_tD_*^\beta} \, u(x;t) =
     {_tD^\beta} \, u(x;t) - \frac{t^{-\beta}}{\Gamma(1-\beta)} \,
     u(x;0) \,, 
\ee 
and (\ref{STFDE0}) becomes 
\be 
\hspace{-0.3truecm}     
{_tD^\beta} \, u(x;t) = {_xD_\theta^\alpha} \, u(x;t) +
     \frac{t^{-\beta}}{\Gamma(1-\beta)} \, u(x;0) \,, 
\,\, -\infty < x < + \infty \,, 
\,\, t \ge 0 \,.
\label{STFDE0-RL}
\ee 

Equation (\ref{STFDE0}) is stated also as 
\be \frac{\partial u}{\partial t} = {_tD^{1-\beta}} \left[{_xD_\theta^\alpha} \,
  u(x;t)\right] \,.
\label{STFDE0-RL2}
\ee 
However, it is possible to show that the fundamental solutions of
(\ref{STFDE0}), (\ref{STFDE0-RL}) and (\ref{STFDE0-RL2}) are equal
\cite{gorenflo_etal-klm-2012}.

When $1<\beta \le 2$ a second initial condition is needed corresponding to 
$\displaystyle{u_t(x;0)=\left.\frac{\partial u}{\partial t}\right|_{t=0}}$,
and two Green functions follow
according to the initial conditions $\{u(x;0)=\delta(x) \,,
u_t(x;0)=0\}$ and $\{u(x;0)=0 \,, u_t(x;0)=\delta(x)\}$, respectively.
However, this second Green function emerges to be a primitive (with
respect to the variable $t$) of the first Green function, so that it
cannot be interpreted as a PDF because it is no
longer normalized over $x$ \cite{mainardi_etal-amc-2003}.  Hence, solely the
first Green function can be considered for diffusion problems.

\smallskip
In general, the fundamental solution $\Kabt(x;t)$ algebraically
decreases as $|x|^{-(\alpha +1)}$, thus it belongs to the domain of
attraction of the L\'evy stable densities of index $\alpha$.
Moreover, $\Kabt(x;t)$ obeys the following self-similarity, or scaling,
relationship: 
\be 
\Kabt(x;t) =
t^{-\beta/\alpha} \, \Kabt\left(\frac{x}{t^{\beta/\alpha}}\right) \,,
\ee 
and it meets the following {\it symmetry relation} 
\be 
\Kabt(-x;t)
= K_{\alpha,\beta}^{-\theta}(x;t) \,,
\label{symmetry}
\ee 
which allows for the restriction of the analysis to $x\ge 0$.  
When $x \ge 0$ the analytical solution of (\ref{STFDE0}) can be expressed
by the following Mellin--Barnes integral representation
\cite{pagnini-tesi-2000,mainardi_etal-fcaa-2001} 
\be  
\hspace{-0.5truecm}
\Kabt(x;t) =
\frac{1}{\alpha x} \frac{1}{2 \pi i}
\int_{\gamma-i\infty}^{\gamma+i\infty}
\frac{\Gamma\left(\frac{s}{\alpha}\right) \,
  \Gamma\left(1-\frac{s}{\alpha}\right) \, \Gamma(1-s)}
     {\Gamma\left(1-\frac{\beta}{\alpha}s\right) \, \Gamma ( \rho \,
       s) \, \Gamma (1-\rho \,s)} \,
     \left(\frac{x}{t^{\beta/\alpha}}\right)^s \, d s \,, 
\hspace{-0.2truecm}
\label{MB}
\ee 
where $\displaystyle{\rho = \frac{\alpha -\theta}{2 \,\alpha}}$
and $\gamma$ is a suitable real constant.  
Solution (\ref{MB}) can be also expressed in terms of H-Fox function
\cite{pagnini-tesi-2000,mainardi_etal-jcam-2005}.

\smallskip
The special cases of space-time fractional diffusion equation (\ref{STFDE0}) 
are the following.

\smallskip
The {\it space-fractional diffusion} equation is obtained when $\{0<\alpha <2 \,,
\beta =1\}$ such that 
\be K_{\alpha,1}^\theta (x;t) = \Lat(x;t) =
t^{-1/\alpha} \Lat\left(\frac{x}{t^{1/\alpha}}\right) \,, \quad 0 < x
< \infty \,, 
\label{levy_flights}
\ee 
where $\Lat(x)$ is the class of strictly stable
densities with algebraic tail decaing as $|x|^{-(\alpha +1)}$ and
infinite variance, where $\alpha$ and  $\theta$ are the scaling and asymmetry 
parameters, respectively.  Moreover, stable PDFs with $0 < \alpha < 1$ and
extremal value of the asymmetry parameter $\theta$ are one-sided with
support $R_0^+$ if $\theta =-\alpha$ and $R_0^-$ if $\theta=+\alpha$.

\smallskip
The {\it time-fractional diffusion} equation is obtained when $\{\alpha =2, 0<
\beta <2\}$ such that \be K_{2,\beta }^0 (x;t) = \frac{1}{2} \,
M_{\beta/2}(x;t) = \frac{1}{2} \, t^{-\beta/2}
M_{\beta/2}\l(\frac{x}{t^{\beta/2}}\r) \,, \quad x \ge 0 \,,
\label{Mfunction}
\ee where $M_\nu(x)$, $0 < \nu < 1$, is the M-Wright/Mainardi density
\cite{mainardi_etal-ijde-2010,mainardi_etal-lnsim-2010,
cahoy-fejts-2011,cahoy-cstm-2012,cahoy-cs-2012,pagnini-fcaa-2013,pagnini_etal-caim-2014}
which has stretched exponential tails and finite variance growing in time with the power law $t^\beta$.  
Since $\alpha=2$, according to
(\ref{def:alphabetatheta}), it holds $\theta=0$, then the PDF is
symmetric and the extension to $-\infty < x < + \infty$ is obtained by
replacing $x$ with $|x|$ in (\ref{Mfunction}).

\smallskip
The {\it neutral fractional diffusion} equation is obtained when $\{0<\alpha =
\beta <2\}$ whose solution can be expressed in explicit form by
non-negative simple elementary functions
\cite{saichev_etal-c-1997,mainardi_etal-fcaa-2001}, i.e., when $x \ge 0$
\be 
\hspace{-0.3truecm}
K_{\alpha,\alpha}^{\theta}(x;t) = N_\alpha^\theta(x;t) =
\frac{t^{-1}}{\pi} \, \frac{(x/t)^{\alpha-1} \sin[{\pi\over 2}(\alpha
    -\theta )]} {1 + 2(x/t)^\alpha \cos[{\pi\over 2}(\alpha -\theta)]
  + (x/t)^{2\alpha}} \,.
\label{ndiff}
\ee 

Recently Luchko \cite{luchko-jmp-2013} has considered and analyzed
the case $1 < \alpha < 2$ and $\theta=0$ of (\ref{ndiff}).  
Moreover, the PDF given in equation (\ref{ndiff}), with $0< \alpha < 1$, 
emerged in the study of finite Larmor radius effects on non-diffusive tracer transport 
in a zonal flow \cite{gustafson_etal-pp-2008}. 
Numerical evidences of the same PDF also emerged in non-diffusive chaotic transport by Rossby 
waves in zonal flow \cite{delcastillonegrete-npg-2010}.

\smallskip
The local {\it classical diffusion} equation is obtained when  
$\{\alpha =2, \beta =1\}$, and its Gaussian solution is recovered as a limiting 
case from both the space-fractional ($\alpha =2$) and the time-fractional 
($\beta =1$) diffusion equations, i.e., 
\be 
K_{2,1}^0(x;t) = L_2^0(x;t) =
\frac{1}{2} \, M_{1/2}(x;t) = \mG(x;t)= \frac{\e^{-x^2/(4 t)}}{\sqrt{4 \pi t}} \,, 
\,\,\,  x \ge 0 \,.  
\label{gaussiancase}
\ee

\smallskip
The last special case is the limit case of the {\it D'Alembert wave
equation}, $\{\alpha =2, \beta =2\}$, and it holds 
\be K_{2,2}^0(x;t)
= \frac{1}{2} \, M_1(x;t) = \frac{1}{2} \, \delta(x-t) \,, \quad 0 < x
< \infty \,.  
\ee

\section*{Acknowledgements}
This research is supported by MINECO under Grant MTM2013-40824-P, 
by Bizkaia Talent and European Commission through COFUND programme under Grant AYD-000-252,
and also by the Basque Government through the BERC 2014-2017 program 
and by the Spanish Ministry of Economy and Competitiveness MINECO: BCAM Severo Ochoa accreditation SEV-2013-0323. 



\begin{thebibliography}{150}
 \normalsize

\bibitem{akin_jsmte09}
O.~C. Akin, P.~Paradisi, P.~Grigolini,
\newblock Perturbation-induced emergence of poisson-like behavior in
  non-poisson systems.
\newblock {\em J. Stat. Mech.: Theory Exp.}, (2009), P01013.

\bibitem{akin_pa06}
O.C. Akin, P.~Paradisi, P.~Grigolini,
\newblock Periodic trend and fluctuations: The case of strong correlation.
\newblock {\em Physica A} {\bf 371}, (2006), 157--170.

\bibitem{allegrini_csf13}
P.~Allegrini, P.~Paradisi, D.~Menicucci, M.~Laurino, R.~Bedini, A.~Piarulli, A.~Gemignani,
\newblock Sleep unconsciousness and breakdown of serial critical intermittency: New vistas on the global workspace.
\newblock {\em Chaos Solitons Fract.} {\bf 55}, (2013), 32--43.

\bibitem{baeumer_etal-fcaa-2001}
B.~Baeumer, M.~M. Meerschaert,
\newblock Stochastic solutions for fractional {Cauchy} problems.
\newblock {\em Fract. Calc. Appl. Anal.} {\bf 4}, (2001), 481--500.

\bibitem{baeumer_etal-tams-2009}
B.~Baeumer, M.~M. Meerschaert, E.~Nane,
\newblock Brownian subordinators and fractional {Cauchy} problems.
\newblock {\em T. Am. Math. Soc.} {\bf 361}, No 7 (2009), 3915--3930.

\bibitem{baeumer_etal-jap-2009}
B.~Baeumer, M.~M. Meerschaert, E.~Nane,
\newblock Space-time fractional diffusion.
\newblock {\em J. Appl. Prob.} {\bf 46}, (2009), 1100--1115.

\bibitem{baleanu_etal-2012}
D.~Baleanu, K.~Diethelm, E.~Scalas, J.~J. Trujillo,
\newblock {\em Fractional Calculus: Models and Numerical Methods}.
\newblock World Scientific Publishers, New Jersey (2012).
\newblock Series on Complexity, Nonlinearity and Chaos, volume 3.

\bibitem{barkai-cp-2002}
E.~Barkai,
\newblock {CTRW} pathways to the fractional diffusion equation.
\newblock {\em Chem. Phys.} {\bf 284}, (2002), 13--27.

\bibitem{benson_etal-awr-2013}
D.~A. Benson, M.~M. Meerschaert, J.~Revielle,
\newblock Fractional calculus in hydrologic modeling: {A} numerical perspective.
\newblock {\em Adv. Water Resour.} {\bf 51}, (2013), 479--497.

\bibitem{biagini_etal-2008}
F.~Biagini, Y.~Hu, B.~{\O}ksendal, T.~Zhang,
\newblock {\em Stochastic Calculus for Fractional Brownian Motion and Applications}.
\newblock Springer (2008).

\bibitem{bianco_cpl07}
S.~Bianco, P.~Grigolini, P.~Paradisi,
\newblock A fluctuating environment as a source of periodic modulation.
\newblock {\em Chem. Phys. Lett.} {\bf 438}, No 4-6 (2007), 336--340.

\bibitem{cahoy-fejts-2011}
D.~O. Cahoy,
\newblock On the parametrization of the {M-Wright} function.
\newblock {\em Far East J. Theor. Stat.} {\bf 34}, No 2 (2011), 155--164.

\bibitem{cahoy-cstm-2012}
D.~O. Cahoy,
\newblock Estimation and simulation for the {M-Wright} function.
\newblock {\em Commun. Stat.-Theor. M.} {\bf 41}, No 8 (2012), 1466--1477.

\bibitem{cahoy-cs-2012}
D.~O. Cahoy,
\newblock Moment estimators for the two-parameter {M-Wright} distribution.
\newblock {\em Computation. Stat.} {\bf 27}, No 3 (2012), 487--497.

\bibitem{castiglione_etal-pd-1999}
P.~Castiglione, A.~Mazzino, P.~Muratore-Ginanneschi, A.~Vulpiani,
\newblock On strong anomalous diffusion.
\newblock {\em Physica D} {\bf 134}, (1999), 75--93.

\bibitem{chambers_etal-jasa-1976}
J.~M. Chambers, C.~L. Mallows, B.~W. Stuck,
\newblock A method for simulating skewed stable random variables.
\newblock {\em J. Amer. Statist. Assoc.} {\bf 71}, (1976), 340--344.

\bibitem{chevrollier_etal-epjd-2010}
M.~Chevrollier, N.~Mercadier, W.~Guerin, R.~Kaiser,
\newblock Anomalous photon diffusion in atomic vapors.
\newblock {\em Eur. Phys. J. D} {\bf 58}, (2010), 161--165.

\bibitem{compte-pre-1996}
A.~Compte,
\newblock Stochastic foundations of fractional dynamics.
\newblock {\em Phys. Rev. E} {\bf 53}, No 4 (1996), 4191--4193.

\bibitem{cox_1962}
D.R. Cox,
\newblock {\em Renewal Theory}.
\newblock Methuen \& Co. Ltd., London (1962).

\bibitem{dasilva-ijmpcs-2015}
J.~L. da~Silva,
\newblock Local times for {grey Brownian motion}.
\newblock {\em Int. J. Mod. Phys. Conf. Ser.} {\bf 36}, (2015), 1560003.
\newblock [7th Jagna International Workshop (2014)].

\bibitem{dasilva_etal-s-2014}
J.~L. da~Silva, M.~Erraoui,
\newblock {Grey Brownian motion} local time: {Existence} and
  weak-approximation.
\newblock {\em Stochastics} {\bf 87}, (2014), 347--361.

\bibitem{delcastillonegrete-pp-2004}
D.~del Castillo-Negrete,
\newblock Fractional diffusion in plasma turbulence.
\newblock {\em Phys. Plasmas} {\bf 11}, No 8 (2004), 3854--3864.

\bibitem{delcastillonegrete-npg-2010}
D.~del Castillo-Negrete,
\newblock Non-diffusive, non-local transport in fluids and plasmas.
\newblock {\em Nonlin. Processes Geophys.} {\bf 17}, (2010), 795--807.

\bibitem{delcastillonegrete_etal-prl-2005}
D.~del Castillo-Negrete, B.~A. Carreras, V.~E. Lynch,
\newblock Nondiffusive transport in plasma turbulence: {A} fractional diffusion
  approach.
\newblock {\em Phys. Rev. Lett.} {\bf 94}, (2005), 065003.

\bibitem{delcastillonegrete_etal-nf-2008}
D.~del Castillo-Negrete, P.~Mantica, V.~Naulin, J.~J. Rasmussen, JET~EFDA
  contributors,
\newblock Fractional diffusion models of non-local perturbative transport:
  numerical results and application to {JET} experiments.
\newblock {\em Nucl. Fusion} {\bf 48}, (2008), 075009.

\bibitem{dieker-phd-2004}
T.~Dieker,
\newblock {\em Simulation of Fractional Brownian Motion}.
\newblock CWI and University of Twente, The Netherlands (2004).
\newblock Ph.D. Thesis, Department of Mathematical Sciences, University of
  Twente, The Netherlands.

\bibitem{dieterich_etal-pnas-2008}
P.~Dieterich, R.~Klages, R.~Preuss, A.~Schwab,
\newblock Anomalous dynamics of cell migration.
\newblock {\em Proc. Nat. Acad. Sci.} {\bf 105}, No 2 (2008), 459--463.

\bibitem{difpradalier_etal-pre-2010}
G.~Dif-Pradalier, P.~H. Diamond, V.~Grandgirard, Y.~Sarazin, J.~Abiteboul,
  X.~Garbet, Ph. Ghendrih, A.~Strugarek, S.~Ku, C.~S. Chang,
\newblock On the validity of the local diffusive paradigm in turbulent plasma
  transport.
\newblock {\em Phys. Rev. E} {\bf 82}, (2010), 025401(R).

\bibitem{dybiec-jsmte-2009}
B.~Dybiec,
\newblock Anomalous diffusion: temporal {non-Markovianity} and weak ergodicity
  breaking.
\newblock {\em J. Stat. Mech.-Theory Exp.}, (2009), P08025.

\bibitem{dybiec_etal-c-2010}
B.~Dybiec, E.~Gudowska-Nowak,
\newblock Subordinated diffusion and continuous time random walk asymptotics.
\newblock {\em Chaos} {\bf 20}, No 4 (2010), 043129.

\bibitem{eule_etal-epl-2009}
S.~Eule, R.~Friedrich,
\newblock Subordinated {Langevin} equations for anomalous diffusion in external
  potentials--biasing and decoupled external forces.
\newblock {\em Europhys. Lett.} {\bf 86}, (2009), 3008.

\bibitem{feller-1971}
W.~Feller,
\newblock {\em An Introduction to Probability Theory and its Applications},
  volume~2.
\newblock Wiley, New York (1971), second edition.

\bibitem{fogedby-pre-1994}
H.~C. Fogedby,
\newblock Langevin equations for continuous time {L\'evy} flights.
\newblock {\em Phys. Rev. E} {\bf 50}, No 2 (1994), 1657--1660.

\bibitem{fulger_etal-pre-2008}
D.~Fulger, E.~Scalas, G.~Germano,
\newblock Monte {Carlo} simulation of uncoupled continuous-time random walks
  yielding a stochastic solution of the space-time fractional diffusion
  equation.
\newblock {\em Phys. Rev. E} {\bf 77}, (2008), 021122.

\bibitem{fulger_etal-fcaa-2013}
D.~Fulger, E.~Scalas, G.~Germano,
\newblock Random numbers form the tails of probability distributions using the
  transformation method.
\newblock {\em Fract. Calc. Appl. Anal.} {\bf 16}, No 2 (2013), 332--353.

\bibitem{germano_etal-pre-2009}
G.~Germano, M.~Politi, E.~Scalas, R.~L. Schilling,
\newblock Stochastic calculus for uncoupled continuous-time random walks.
\newblock {\em Phys. Rev. E} {\bf 79}, No 6 (2009), 066102.

\bibitem{gorenflo_etal-fcaa-2000}
R.~Gorenflo, A.~Iskenderov, {Yu.} Luchko,
\newblock Mapping between solutions of fractional diffusion-wave equations.
\newblock {\em Fract. Calc. Appl. Anal.} {\bf 3}, (2000), 75--86.

\bibitem{gorenflo_etal-fcaa-1998}
R.~Gorenflo, F.~Mainardi,
\newblock Random walk models for space-fractional diffusion processes.
\newblock {\em Fract. Calc. Appl. Anal.} {\bf 1}, No 2 (1998), 167--191.

\bibitem{gorenflo_etal-jcam-2009}
R.~Gorenflo, F.~Mainardi,
\newblock Some recent advances in theory and simulation of fractional diffusion
  processes.
\newblock {\em J. Comput. Appl. Math.} {\bf 229}, No 2 (2009), 400--415.

\bibitem{gorenflo_etal-epjst-2011}
R.~Gorenflo, F.~Mainardi,
\newblock Subordination pathways to fractional diffusion.
\newblock {\em Eur. Phys. J. Special Topics} {\bf 193}, (2011), 119--132.

\bibitem{gorenflo_etal-klm-2012}
R.~Gorenflo, F.~Mainardi,
\newblock Parametric subordination in fractional diffusion processes.
\newblock In: J.~Klafter, S.~C. Lim, and R.~Metzler, editors, {\em Fractional
  Dynamics. Recent Advances}, World Scientific, Singapore (2012), 227--261.

\bibitem{gorenflo_etal-cp-2002}
R.~Gorenflo, F.~Mainardi, D.~Moretti, G.~Pagnini, P.~Paradisi,
\newblock Discrete random walk models for space-time fractional diffusion.
\newblock {\em Chem. Phys.} {\bf 284}, (2002), 521--541.

\bibitem{gorenflo_etal-pa-2002}
R.~Gorenflo, F.~Mainardi, D.~Moretti, G.~Pagnini, P.~Paradisi,
\newblock Fractional diffusion: probability distributions and random walk
  models.
\newblock {\em Physica A} {\bf 305}, No 1-2 (2002), 106--112.

\bibitem{gorenflo_etal-nd-2002}
R.~Gorenflo, F.~Mainardi, D.~Moretti, P.~Paradisi,
\newblock Time fractional diffusion: {A} discrete random walk approach.
\newblock {\em Nonlinear Dynam.} {\bf 29}, No 1-4 (2002), 129--143.

\bibitem{gorenflo_etal-csf-2007}
R.~Gorenflo, F.~Mainardi, A.~Vivoli,
\newblock Continuous-time random walk and parametric subordination in
  fractional diffusion.
\newblock {\em Chaos Solitons Fract.} {\bf 34}, No 1 (2007), 87--103.

\bibitem{grigolini_etal-pre-1999}
P.~Grigolini, A.~Rocco, B.~J. West,
\newblock Fractional calculus as a macroscopic manifestation of randomness.
\newblock {\em Phys. Rev. E} {\bf 59}, No 3 (1999), 2603--2613.

\bibitem{gustafson_etal-pp-2008}
K.~Gustafson, D.~del Castillo-Negrete, W.~Dorland,
\newblock Finite {Larmor} radius effects on nondiffusive tracer transport in
  zonal flows.
\newblock {\em Phys. Plasmas} {\bf 15}, (2008), 102309.

\bibitem{honkonen-pre-1996}
J.~Honkonen,
\newblock Stochastic processes with stable distributions in random
  environments.
\newblock {\em Phys. Rev. E} {\bf 55}, No 1 (1996), 327--331.

\bibitem{hosking-wrr-1984}
J.~R.~M. Hosking,
\newblock Modeling persistence in hydrological time series using fractional
  differencing.
\newblock {\em Water Resour. Res.} {\bf 20}, (1984), 1898--1908.

\bibitem{hughes-pre-2002}
B.~D. Hughes,
\newblock Anomalous diffusion, stable processes, and generalized functions.
\newblock {\em Phys. Rev. E} {\bf 65}, (2002), 035105(R).

\bibitem{klafter_etal-pw-2005}
J.~Klafter, I.~M. Sokolov,
\newblock Anomalous diffusion spread its wings.
\newblock {\em Physics World} {\bf 18}, (2005), 29--32.

\bibitem{kleinhans_etal-pre-2007}
D.~Kleinhans, R.~Friedrich,
\newblock Continuous-time random walks: {Simulation} of continuous
  trajectories.
\newblock {\em Phys. Rev. E} {\bf 76}, (2007), 061102.

\bibitem{leoncini_etal-csf-2004}
X.~Leoncini, L.~Kuznetsov, G.~M. Zaslavsky,
\newblock Evidence of fractional transport in point vortex flow.
\newblock {\em Chaos Solitons Fract.} {\bf 19}, (2004), 259--273.

\bibitem{luchko-jmp-2013}
{Yu.} Luchko,
\newblock Fractional wave equation and damped waves.
\newblock {\em J. Math. Phys.} {\bf 54}, (2013), 031505.

\bibitem{magdziarz_etal-prl-2008}
M.~Magdziarz, A.~Weron, J.~Klafter,
\newblock Equivalence of the fractional {Fokker--Planck} and subordinated
  {Langevin} equations: {The} case of a time-dependent force.
\newblock {\em Phys. Rev. Lett.} {\bf 101}, (2008), 210601.

\bibitem{mainardi-csf-1996}
F.~Mainardi,
\newblock Fractional relaxation-oscillation and fractional diffusion-wave
  phenomena.
\newblock {\em Chaos Solitons Fract.} {\bf 7}, (1996), 1461--1477.

\bibitem{mainardi_etal-fcaa-2001}
F.~Mainardi, Yu. Luchko, G.~Pagnini,
\newblock The fundamental solution of the space-time fractional diffusion
  equation.
\newblock {\em Fract. Calc. Appl. Anal.} {\bf 4}, No 2 (2001), 153--192.

\bibitem{mainardi_etal-lnsim-2010}
F.~Mainardi, A.~Mura, G.~Pagnini,
\newblock The functions of the {Wright} type in fractional calculus.
\newblock {\em Lecture Notes of Seminario Interdisciplinare di Matematica} {\bf 9}, (2010), 111--128.

\bibitem{mainardi_etal-ijde-2010}
F.~Mainardi, A.~Mura, G.~Pagnini,
\newblock The {M-Wright} function in time-fractional diffusion processes: {A}
  tutorial survey.
\newblock {\em Int. J. Differ. Equations} {\bf 2010}, (2010), 104505.

\bibitem{mainardi_etal-amc-2003}
F.~Mainardi, G.~Pagnini,
\newblock The {Wright} functions as solutions of the time-fractional diffusion
  equations.
\newblock {\em Appl. Math. Comput.} {\bf 141}, (2003), 51--62.

\bibitem{mainardi_etal-fcaa-2003}
F.~Mainardi, G.~Pagnini, R.~Gorenflo,
\newblock Mellin transform and subordination laws in fractional diffusion
  processes.
\newblock {\em Fract. Calc. Appl. Anal.} {\bf 6}, No 4 (2003), 441--459.

\bibitem{mainardi_etal-jms-2006}
F.~Mainardi, G.~Pagnini, R.~Gorenflo,
\newblock Mellin convolution for subordinated stable processes.
\newblock {\em J. Math. Sci.} {\bf 132(5)}, (2006), 637--642.

\bibitem{mainardi_etal-jcam-2005}
F.~Mainardi, G.~Pagnini, R.~K. Saxena,
\newblock Fox {H} functions in fractional diffusion.
\newblock {\em J. Comput. Appl. Math.} {\bf 178}, (2005), 321--331.

\bibitem{meerschaert_etal-pre-2002}
M.~M. Meerschaert, D.~A. Benson, H.-P. Scheffler, B.~Baeumer,
\newblock Stochastic solution of space-time fractional diffusion equations.
\newblock {\em Phys. Rev. E} {\bf 65}, (2002), 041103.

\bibitem{meerschaert_etal-2012}
M.~M. Meerschaert, A.~Sikorskii,
\newblock {\em Stochastic Models for Fractional Calculus}.
\newblock De Gruyter (2012).

\bibitem{meroz_etal-prl-2011}
Y. Meroz, I.~M. Sokolov, J.~Klafter,
\newblock Unequal twins: Probability distributions do not determine everything.
\newblock {\em Phys. Rev. Lett.} {\bf 107}, (2011), 260601.

\bibitem{metzler_etal-jpa-2004}
R.~Metzler, J.~Klafter,
\newblock The restaurant at the end of the random walk: recent developments in
  fractional dynamics descriptions of anomalous dynamical processes.
\newblock {\em J. Phys. A: Math. Theor.} {\bf 37}, No 31 (2004), R161--R208.

\bibitem{metzler_etal-cp-2002}
R.~Metzler, T.~F. Nonnenmacher,
\newblock Space- and time-fractional diffusion and wave equations, fractional
  Fokker--Planck equations, and physical motivation.
\newblock {\em Chem. Phys.} {\bf 284}, (2002), 67--90.

\bibitem{montroll1964}
E.~W. Montroll,
\newblock Random walks on lattices.
\newblock {\em Proc. Symp. Appl. Math. Am. Math. Soc.} {\bf 16}, (1964), 193--220.

\bibitem{montroll_etal-jmp-1965}
E.~W. Montroll, G.~H. Weiss,
\newblock Random walks on lattices. {II}.
\newblock {\em J. Math. Phys.} {\bf 6}, (1965), 167--181.

\bibitem{mura-phd-2008}
A.~Mura,
\newblock {\em Non-Markovian Stochastic Processes and Their Applications: From
  Anomalous Diffusion to Time Series Analysis}.
\newblock Lambert Academic Publishing (2011).
\newblock Ph.D. Thesis, Physics Department, University of Bologna (2008).

\bibitem{mura_etal-itsf-2009}
A.~Mura, F.~Mainardi,
\newblock A class of self-similar stochastic processes with stationary
  increments to model anomalous diffusion in physics.
\newblock {\em Integr. Transf. Spec. F.} {\bf 20}, No 3-4 (2009), 185--198.

\bibitem{mura_etal-jpa-2008}
A.~Mura, G.~Pagnini,
\newblock Characterizations and simulations of a class of stochastic processes
  to model anomalous diffusion.
\newblock {\em J. Phys. A: Math. Theor.} {\bf 41}, (2008), 285003.

\bibitem{pagnini-fcaa-2012}
G.~Pagnini,
\newblock Erd\'elyi--{Kober} fractional diffusion.
\newblock {\em Fract. Calc. Appl. Anal.} {\bf 15}, No 1 (2012), 117--127.

\bibitem{pagnini-fcaa-2013}
G.~Pagnini,
\newblock The {M-Wright} function as a generalization of the {Gaussian} density
  for fractional diffusion processes.
\newblock {\em Fract. Calc. Appl. Anal.} {\bf 16}, No 2 (2013), 436--453.

\bibitem{pagnini-mesa-2014}
G.~Pagnini,
\newblock Self-similar stochastic models with stationary increments for
  symmetric space-time fractional diffusion.
\newblock In: {\em Proceedings of the 10th IEEE/ASME International Conference on
  Mechatronic and Embedded Systems and Applications, MESA 2014}, 
Senigallia (AN), Italy, 10--12 September (2014), Paper Code MESA2014 003. 
Print ISBN:978-1-4799-2772-2; INSPEC Accession Number:14701095; doi:10.1109/MESA.2014.6935520.

\bibitem{pagnini-pa-2014}
G.~Pagnini,
\newblock Short note on the emergence of fractional kinetics.
\newblock {\em Physica A} {\bf 409}, (2014), 29--34.

\bibitem{pagnini-pmam-2014}
G.~Pagnini,
\newblock Subordination formulae for space-time fractional diffusion processes
  via {Mellin} convolution.
\newblock In: P.~M. Pardalos, R.~P. Agarwal, L.~Ko{\v{c}}cinac, R.~Neck,
  N.~Mastorakis, and K.~Ntalianas, editors, {\em Recent Advances in
  Mathematics, Statistics and Economics. Proceedings of the 2014 International
  Conference on Pure Mathematics -- Applied Mathematics (PM-AM'14)}, Venice, Italy, 15--17 March (2014), 40--45.
\newblock ISBN: 978-1-61804-225-5.

\bibitem{pagnini-tesi-2000}
G.~Pagnini,
\newblock {\em Generalized Equations for Anomalous Diffusion and their
  Fundamental Solutions}.
\newblock Thesis for Degree in Physics, University of Bologna, October (2000).
\newblock In Italian.

\bibitem{pagnini_etal-ijsa-2012}
G.~Pagnini, A.~Mura, F.~Mainardi,
\newblock Generalized fractional master equation for self-similar stochastic
  processes modelling anomalous diffusion.
\newblock {\em Int. J. Stoch. Anal.} {\bf 2012}, (2012), 427383.

\bibitem{pagnini_etal-ptrsa-2013}
G.~Pagnini, A.~Mura, F.~Mainardi,
\newblock Two-particle anomalous diffusion: {Probability} density functions and
  self-similar stochastic processes.
\newblock {\em Phil. Trans. R. Soc. A} {\bf 371}, (2013), 20120154.

\bibitem{pagnini_etal-caim-2014}
G.~Pagnini, E.~Scalas,
\newblock Historical notes on the {M-Wright/Mainardi} function.
\newblock {\em Communications in Applied and Industrial Mathematics} {\bf 6}, No 1 (2014), e--496.
\newblock DOI: 10.1685/journal.caim.496 (Editorial).

\bibitem{paradisi_caim15}
P.~Paradisi,
\newblock Fractional calculus in statistical physics: The case of time
  fractional diffusion equation.
\newblock {\em Communications in Applied and Industrial Mathematics} {\bf 6}, No 2 (2014), e--530.
\newblock doi: 10.1685/journal.caim.530.

\bibitem{paradisi_aipcp13}
P.~Paradisi, P.~Allegrini, A.~Gemignani, M.~Laurino, D.~Menicucci, A.~Piarulli,
\newblock Scaling and intermittency of brain events as a manifestation of
  consciousness.
\newblock {\em AIP Conf. Proc.} {\bf 1510}, (2013), 151--161.

\bibitem{paradisi_epjst09}
P.~Paradisi, R.~Cesari, D.~Contini, A.~Donateo, L.~Palatella,
\newblock Characterizing memory in atmospheric time series: an alternative
  approach based on renewal theory.
\newblock {\em Eur. Phys. J. Special Topics} {\bf 174}, (2009), 207--218.

\bibitem{paradisi_romp12}
P.~Paradisi, R.~Cesari, A.~Donateo, D.~Contini, P.~Allegrini,
\newblock Diffusion scaling in event-driven random walks: an application to
  turbulence.
\newblock {\em Rep. Math. Phys.} {\bf 70}, (2012), 205--220.

\bibitem{paradisi_npg12}
P.~Paradisi, R.~Cesari, A.~Donateo, D.~Contini, P.~Allegrini,
\newblock Scaling laws of diffusion and time intermittency generated by
  coherent structures in atmospheric turbulence.
\newblock {\em Nonlin. Processes Geophys.} {\bf 19}, (2012), 113--126.
\newblock Corrigendum, {\it Nonlin. Processes Geophys.} {\bf 19}, (2012), 685.

\bibitem{paradisi_cejp09}
P.~Paradisi, R.~Cesari, P.~Grigolini,
\newblock Superstatistics and renewal critical events.
\newblock {\em Cent. Eur. J. Phys.} {\bf 7}, (2009), 421--431.

\bibitem{paradisi_pceb01}
P.~Paradisi, R.~Cesari, F.~Mainardi, A.~Maurizi, F.~Tampieri,
\newblock A generalized {Fick's} law to describe non-local transport effects.
\newblock {\em Phys. Chem. Earth} {\bf 26}, No 4 (2001), 275--279.

\bibitem{paradisi_pa01}
P.~Paradisi, R.~Cesari, F.~Mainardi, F.~Tampieri,
\newblock The fractional {Fick's} law for non-local transport processes.
\newblock {\em Physica A} {\bf 293}, No 1-2 (2001), 130--142.

\bibitem{paradisi_bmcsb15}
P.~Paradisi, D.~Chiarugi, P.~Allegrini,
\newblock A renewal model for the emergence of anomalous solute crowding in
  liposomes.
\newblock {\em BMC Syst. Biol.} {\bf 9}, suppl 3 (2015), s7. 

\bibitem{podlubny-1999}
I.~Podlubny,
\newblock {\em Fractional Differential Equations}.
\newblock Academic Press, San Diego (1999).

\bibitem{ratynskaia_etal-prl-2006}
S.~Ratynskaia, K.~Rypdal, C.~Knapek, S.~Khrapak, A.~V. Milovanov, A.~Ivlev,
  J.~J. Rasmussen, G.~E. Morfill,
\newblock Superdiffusion and viscoelastic vortex flows in a two-dimensional
  complex plasma.
\newblock {\em Phys. Rev. Lett.} {\bf 96}, No 10 (2006), 105010.

\bibitem{rocco_etal-pa-1999}
A.~Rocco, B.~J. West,
\newblock Fractional calculus and the evolution of fractal phenomena.
\newblock {\em Physica A} {\bf 265}, No 3-4 (1999), 535--546.

\bibitem{saichev_etal-c-1997}
A.~Saichev, G.~Zaslavsky,
\newblock Fractional kinetic equations: solutions and applications.
\newblock {\em Chaos} {\bf 7}, (1997), 753--764.

\bibitem{scalas_etal-pa-2000}
E.~Scalas, R.~Gorenflo, F.~Mainardi,
\newblock Fractional calculus and continuous-time finance.
\newblock {\em Physica A} {\bf 284}, (2000), 376--384.

\bibitem{scalas_etal-pre-2004}
E.~Scalas, R.~Gorenflo, F.~Mainardi,
\newblock Uncoupled continuous-time random walks: {Solution} and limiting
  behavior of the master equation.
\newblock {\em Phys. Rev. E} {\bf 69}, (2004), 011107.

\bibitem{schmiedeberg_etal-jsmte-2009}
M.~Schmiedeberg, V.~{Yu.} Zaburdaev, H.~Stark,
\newblock On moments and scaling regimes in anomalous random walks.
\newblock {\em J. Stat. Mech.-Theory Exp.}, (2009), P12020.

\bibitem{schneider_etal-jmp-1989}
W.~R. Schneider, W.~Wyss,
\newblock Fractional diffusion and wave equations.
\newblock {\em J. Math. Phys.} {\bf 30}, No 1 (1989), 134--144.

\bibitem{schulz_etal-jpa-2013}
J. H. P. Schulz, A. V. Chechkin, R. Metzler,
\newblock Correlated continuos time random walks: combining scale-invariance with long-range memory for
spatial and temporal dynamics.
\newblock {\em J. Phys. A: Math. Theor.} {\bf 46}, (2013), 475001.

\bibitem{sokolov_etal-pt-2002}
I.~M. Sokolov, J.~Klafter, A.~Blumen,
\newblock Fractional kinetics.
\newblock {\em Physics Today} {\bf 55}, (2002), 48--54.

\bibitem{sokolov_etal-jpa-2004}
I.~M. Sokolov, R. Metzler,
\newblock Non-uniqueness of the first passage time density of L\'evy random processes.
\newblock {\em J. Phys. A: Math. Theor.} {\bf 37}, (2004), L609--L615.

\bibitem{uchaikin-ijtp-2000}
V.~V. Uchaikin,
\newblock {Montroll--Weiss} problem, fractional equations and stable
  distributions.
\newblock {\em Int. J. Theor. Phys.} {\bf 39}, (2000), 2087--2105.

\bibitem{uchaikin_etal-1999}
V.~V. Uchaikin, V.~M. Zolotarev,
\newblock {\em Chance and Stability. Stable Distributions and their
  Applications}.
\newblock VSP, Utrecht (1999).

\bibitem{weiss1983}
G.~H. Weiss, R.~J. Rubin,
\newblock Random walks: Theory and selected applications.
\newblock {\em Adv. Chem. Phys.} {\bf 52}, (1983), 363--505.

\bibitem{weron_etal-pre-2008}
A.~Weron, M.~Magdziarz, K.~Weron,
\newblock Modeling of subdiffusion in space-time-dependent force fields beyond
  the fractional {Fokker--Planck} equation.
\newblock {\em Phys. Rev. E} {\bf 77}, (2008), 036704.

\bibitem{weron-spl-1996}
R.~Weron,
\newblock On the {Chambers--Mallows--Stuck} method for simulating skewed stable
  random variables.
\newblock {\em Statist. Probab. Lett.} {\bf 28}, (1996), 165--171.
\newblock Corrigendum: {\tt
  http://mpra.ub.uni-muenchen.de/20761/1/RWeron96\_Corr.pdf} or {\tt
  http://www.im.pwr.wroc.pl/$\sim$hugo/RePEc/wuu/wpaper/HSC\_96\_01.pdf}.

\bibitem{zaslavsky-1992}
G.~M. Zaslavsky,
\newblock Anomalous transport and fractal kinetics.
\newblock In: H.~K. Moffatt, G.~M. Zaslavsky, P.~Compte, and M.~Tabor, editors,
  {\em Topological Aspects of the Dynamics of Fluids and Plasmas}, Kluwer, Dordrecht (1992), 481--491.
\newblock NATO ASI Series, volume 218.

\bibitem{zaslavsky-pd-1994}
G.~M. Zaslavsky,
\newblock Fractional kinetic equation for {Hamiltonian} chaos.
\newblock {\em Physica D} {\bf 76}, (1994), 110--122.

\bibitem{zaslavsky-c-1994}
G.~M. Zaslavsky,
\newblock Renormalization group theory of anomalous transport in systems with
  {Hamiltonian} chaos.
\newblock {\em Chaos} {\bf 4}, (1994), 25--33.

\bibitem{zaslavsky-pr-2002}
G.~M. Zaslavsky,
\newblock Chaos, fractional kinetics, and anomalous transport.
\newblock {\em Phys. Rep.} {\bf 371}, (2002), 461--580.

\bibitem{zaslavsky-2005}
G.~M. Zaslavsky,
\newblock {\em Hamiltonian Chaos and Fractional Dynamics}.
\newblock Oxford University Press (2005).

\bibitem{zaslavsky_etal-pr-1997}
G.~M. Zaslavsky, B.~A. Niyazov,
\newblock Fractional kinetics and accelerator modes.
\newblock {\em Phys. Rep.} {\bf 283}, (1997), 73--93.

\end{thebibliography}


 \bigskip \smallskip

 \it

 \noindent
$^1$ BCAM -- Basque Center for Applied Mathematics \\
Alameda de Mazarredo 14, E--48009 Bilbao, Basque Country -- SPAIN \\

 \noindent
$^2$ Ikerbasque - Basque Foundation for Science \\
Calle de Mar\'ia D\'iaz de Haro 3, E--48013 Bilbao, Basque Country -- SPAIN \\[4pt]
e-mail: gpagnini@bcamath.org
  
\hfill \\[12pt]
$^3$ ISTI--CNR \\
Istituto di Scienza e Tecnologie dell'Informazione "A. Faedo"\\
Via Moruzzi 1, I--56124 Pisa, ITALY \\[4pt]
e-mail: paolo.paradisi@isti.cnr.it

\end{document}